\documentclass[aps,prc,superscriptaddress,showpacs,amssymb,amsmath,amsfonts,twocolumn,floatfix]{revtex4}
\usepackage{graphicx}
 \usepackage{multirow} 
\begin{document}

\newcommand{\piZeroRxn}{\gamma \ p \rightarrow p \ \pi^0}
\newcommand{\xRxn}{\gamma \ p \rightarrow p \ X}
\newcommand{\ppipiRxn}{\gamma \ p \rightarrow p \ \pi^- \ \pi^+}
\newcommand{\ppipiyRxn}{\gamma \ p \rightarrow p \ \pi^- \ \pi^+ \ Y}
\newcommand{\Eg}{E_{\gamma}}
\newcommand{\cosThetaCm}{\cos\theta_{\rm c.m.}}
\newcommand{\cosThetaCmP}{\cos\theta^p_{\rm c.m.}}
\newcommand{\cosThetaCmPi}{\cos\theta^{\pi}_{\rm c.m.}}

\title{$\pi^0$ photoproduction on the proton for photon energies from
0.675 to 2.875~GeV}

\newcommand*{\ASU}{Arizona State University, Tempe, Arizona 85287-1504}
\affiliation{\ASU}
\newcommand*{\GWU}{The George Washington University, Washington, DC 20052}
\affiliation{\GWU}
\newcommand*{\UCLA}{University of California at Los Angeles, Los Angeles, California  90095-1547}
\affiliation{\UCLA}
\newcommand*{\CSUDH}{California State University, Dominguez Hills, Carson, California 90747-0005}
\affiliation{\CSUDH}
\newcommand*{\CMU}{Carnegie Mellon University, Pittsburgh, Pennsylvania 15213}
\affiliation{\CMU}
\newcommand*{\CUA}{Catholic University of America, Washington, DC 20064}
\affiliation{\CUA}
\newcommand*{\SACLAY}{CEA-Saclay, Service de Physique Nucl\'eaire, F91191 Gif-sur-Yvette, France}
\affiliation{\SACLAY}
\newcommand*{\CNU}{Christopher Newport University, Newport News, Virginia 23606}
\affiliation{\CNU}
\newcommand*{\UCONN}{University of Connecticut, Storrs, Connecticut 06269}
\affiliation{\UCONN}
\newcommand*{\DUKE}{Duke University, Durham, North Carolina 27708-0305}
\affiliation{\DUKE}
\newcommand*{\ECOSSEE}{Edinburgh University, Edinburgh EH9 3JZ, United Kingdom}
\affiliation{\ECOSSEE}
\newcommand*{\FIU}{Florida International University, Miami, Florida 33199}
\affiliation{\FIU}
\newcommand*{\FSU}{Florida State University, Tallahassee, Florida 32306}
\affiliation{\FSU}
\newcommand*{\ECOSSEG}{University of Glasgow, Glasgow G12 8QQ, United Kingdom}
\affiliation{\ECOSSEG}
\newcommand*{\GEISSEN}{Physikalisches Institut der Universitaet Giessen, 35392 Giessen, Germany}
\affiliation{\GEISSEN}
\newcommand*{\ISU}{Idaho State University, Pocatello, Idaho 83209}
\affiliation{\ISU}
\newcommand*{\INFNFR}{INFN, Laboratori Nazionali di Frascati, 00044 Frascati, Italy}
\affiliation{\INFNFR}
\newcommand*{\INFNGE}{INFN, Sezione di Genova, 16146 Genova, Italy}
\affiliation{\INFNGE}
\newcommand*{\ORSAY}{Institut de Physique Nucleaire ORSAY, Orsay, France}
\affiliation{\ORSAY}
\newcommand*{\ITEP}{Institute of Theoretical and Experimental Physics, Moscow, 117259, Russia}
\affiliation{\ITEP}
\newcommand*{\JMU}{James Madison University, Harrisonburg, Virginia 22807}
\affiliation{\JMU}
\newcommand*{\KYUNGPOOK}{Kyungpook National University, 702-701 Daegu, South Korea}
\affiliation{\KYUNGPOOK}
\newcommand*{\MIT}{Massachusetts Institute of Technology, Cambridge, Massachusetts  02139-4307}
\affiliation{\MIT}
\newcommand*{\UMASS}{University of Massachusetts, Amherst, Massachusetts  01003}
\affiliation{\UMASS}
\newcommand*{\MOSCOW}{Moscow State University, General Nuclear Physics Institute, 119899 Moscow, Russia}
\affiliation{\MOSCOW}
\newcommand*{\UNH}{University of New Hampshire, Durham, New Hampshire 03824-3568}
\affiliation{\UNH}
\newcommand*{\NSU}{Norfolk State University, Norfolk, Virginia 23504}
\affiliation{\NSU}
\newcommand*{\OHIOU}{Ohio University, Athens, Ohio  45701}
\affiliation{\OHIOU}
\newcommand*{\ODU}{Old Dominion University, Norfolk, Virginia 23529}
\affiliation{\ODU}
\newcommand*{\PITT}{University of Pittsburgh, Pittsburgh, Pennsylvania 15260}
\affiliation{\PITT}
\newcommand*{\RPI}{Rensselaer Polytechnic Institute, Troy, New York 12180-3590}
\affiliation{\RPI}
\newcommand*{\RICE}{Rice University, Houston, Texas 77005-1892}
\affiliation{\RICE}
\newcommand*{\URICH}{University of Richmond, Richmond, Virginia 23173}
\affiliation{\URICH}
\newcommand*{\SCAROLINA}{University of South Carolina, Columbia, South Carolina 29208}
\affiliation{\SCAROLINA}
\newcommand*{\JLAB}{Thomas Jefferson National Accelerator Facility, Newport News, Virginia 23606}
\affiliation{\JLAB}
\newcommand*{\UNIONC}{Union College, Schenectady, New York 12308}
\affiliation{\UNIONC}
\newcommand*{\VT}{Virginia Polytechnic Institute and State University, Blacksburg, Virginia   24061-0435}
\affiliation{\VT}
\newcommand*{\VIRGINIA}{University of Virginia, Charlottesville, Virginia 22901}
\affiliation{\VIRGINIA}
\newcommand*{\WM}{College of William and Mary, Williamsburg, Virginia 23187-8795}
\affiliation{\WM}
\newcommand*{\YEREVAN}{Yerevan Physics Institute, 375036 Yerevan, Armenia}
\affiliation{\YEREVAN}

\author {M. Dugger}
\affiliation{\ASU}
\author {B.~G.~Ritchie}
\affiliation{\ASU}
\author {J.~P.~Ball}
\affiliation{\ASU}
\author {P. ~Collins}
\affiliation{\ASU}
\author {E.~Pasyuk}
\affiliation{\ASU}
\author {R.~A. Arndt}
\affiliation{\GWU}
\author {W.~J.~Briscoe}
\affiliation{\GWU}
\author {I.~I.~Strakovsky}
\affiliation{\GWU}
\author {R.~L. ~Workman}
\affiliation{\GWU}

\author {G.~Adams}
\affiliation{\RPI}
% ...
\collaboration{The CLAS Collaboration}
     \noaffiliation

\begin{abstract}
Differential cross sections for the reaction $\piZeroRxn$ have been
measured with the CEBAF Large Acceptance Spectrometer (CLAS) and a
tagged photon beam with energies from 0.675 to 2.875~GeV. The results
reported here possess greater accuracy in the absolute normalization
than previous measurements. They disagree with recent CB-ELSA
measurements for the process at forward scattering angles. Agreement
with the SAID and MAID fits is found below 1~GeV. The present set of
cross sections has been incorporated into the SAID database, and
exploratory fits have been extended to 3~GeV. Resonance couplings have
been extracted and compared to previous determinations.
\end{abstract}

\maketitle

\section{Introduction}

The spectrum of baryon resonances has been extensively explored for
clues to the internal structure of nucleons.  Experimental and
phenomenological programs are working in tandem to refine and expand
the known resonance masses, widths, and electromagnetic couplings,
which provide tight constraints for QCD-inspired models, and valuable
benchmarks for lattice calculations.  The most precisely determined
resonance properties are associated with low-lying states and those
higher-mass states with clear Breit-Wigner signatures in amplitudes
where they contribute.  Many other states have been observed, but with
widely-varying mass and width estimates.

Much of the non-strange baryon spectrum has been deduced from fits to
pion-nucleon scattering and photoproduction data.  However there are
difficulties encountered in extracting the states from experimental
data.  While many states have been predicted by QCD-inspired models,
far fewer have been clearly identified.  Some of these missing states
may be weakly coupled to the pion-nucleon channel, and this
possibility has motivated the study of other channels (involving $\eta
N$, $K \Lambda$, and $\pi\pi N$ final states, for example).  Many of
the states established in fits to elastic pion-nucleon scattering data
have $\pi N$ branching ratios~\cite{PDG} only of order 10-20\% and
therefore are not expected to be easily seen in $\pi p$ elastic
scattering.  This is particularly true for resonances with masses
above about 1.7~GeV.
 
Photodecay amplitudes can be extracted from fits to the available
database of charged and neutral pion photoproduction data.  A
knowledge of the resonances contributing to pion-nucleon elastic
scattering is crucial in this task.  Precise measurements of the cross
section and polarization asymmetries are equally important. However,
until very recently, such measurements were lacking in the region
above 1~GeV in photon energy.  Recent Jefferson Lab measurements have
shown that surprises are possible in the behavior of both
polarized~\cite{Cxcz} and unpolarized~\cite{Zhu} cross sections.
Present fits covering photon energies to 2~GeV and beyond are
necessarily model-dependent and underconstrained.

In this paper, we report measurements of the unpolarized differential
cross sections of neutral pion photoproduction on the proton for
incident photon energies from 0.675 to 2.875 GeV.  We have included
the present set of CLAS cross sections in a multipole fit to the
available data covering the resonance region.  At the highest
energies, recent CB-ELSA~\cite{Bonn} measurements are available, and
we have compared both data and fits to show where deviations occur.
Resonance couplings have been extracted for those states that give a
significant contribution to the photoproduction process.  In some
cases, these couplings differ significantly from previous
determinations.

The paper is laid out in the following manner: We give a brief
overview of the experiment in Section~\ref{sec:Exp}.  A more detailed
examination of methods used in the data reduction follows in
Section~\ref{sec:Data}.  The uncertainty estimates for the cross
sections obtained are given in Section~\ref{sec:Errs}.  The
experimental results are described in Section~\ref{sec:Results}.
Various fits to the data are described in Section\ref{sec:fit}, and
the underlying multipole amplitudes and resonance contributions are
displayed and compared to previous determinations in
Section~\ref{sec:ResCoupl}.  Finally, in Section~\ref{sec:conc}, we
provide a brief summary of the results of this study and consider what
extensions of this work would be particularly helpful in the future.

\section{Experiment}
\label{sec:Exp}


The differential cross sections for the reaction $\piZeroRxn$ were
measured with the CEBAF Large Acceptance Spectrometer
(CLAS)~\cite{CLAS} and the bremsstrahlung photon tagging facility
(``photon tagger")~\cite{tag} in Hall B of the Thomas Jefferson
National Accelerator Facility (JLab) as part of a set of experiments
running at the same time with the same experimental configuration
(cryogenic target, tagger, and CLAS) called the ``$g1c$" run period.
The cross sections can be found in electronic form in
Ref. \cite{diffcs} and were part of a program of meson photoproduction
measurements undertaken using the CLAS and photon
tagger~\cite{meson1,meson2,meson3,meson4,meson5,meson6,meson7,meson8}.
The database entries include the differential cross sections, as well
as uncertainties (excluding the overall absolute normalization
uncertainty), for each incident photon energy and $\cosThetaCmPi$
shown in this document.

The full data described here consisted of two running periods with
different incident electron beam energies of 2.445 and 3.115~GeV.
Tagged photons were incident on an 18-cm-long liquid hydrogen target
placed at the center of CLAS.  This target was enclosed by a
scintillator array (start counter) that detected the passage of
charged particles into CLAS from the target~\cite{start}.  The event
trigger required the coincidence of a post-bremsstrahlung electron
passing through the focal plane of the photon tagger and at least one
charged particle detected in CLAS and the start counter.  Tracking of
the charged particles through the magnetic field within CLAS by drift
chambers provided determination of their charge, momentum, and
scattering angle.  This information, together with the particle
velocity measured by the time-of-flight system \cite{TOF} and start
counter, provided particle identification for each particle detected
in CLAS and its corresponding momentum four-vector.  The methods used
for extracting the differential cross sections for $\pi^0$
photoproduction are presented in the next several sections.  The
technique is outlined initially, and then each step is described in
further detail, with data and tests that support the validity of the
approach.

\section{Data reduction}
\label{sec:Data}


Yields for the $\pi^0$ mesons resulting from photoproduction on the
protons within the cryogenic hydrogen target were determined using the
missing mass technique, assuming the two-body reaction $\xRxn$, where
$X$ is the particle hypothesized to be missing.  Briefly, the
following steps are involved in this analysis: \newcounter{bean}
\begin{list}{\arabic{bean}.}
{\usecounter{bean}
\setlength{\rightmargin}
{\leftmargin}}

\item Identify the recoil proton in CLAS, determining the scattering
angle and momentum for the proton.

\item Calculate the missing mass for the recoil proton based on the
assumption that the reaction observed is $\xRxn$.

\item Distribute the events of the resulting missing mass spectra 
into bins defined by incident photon energy $\Eg$ and $\pi^0$
center-of-mass scattering angle $\cosThetaCmPi$, based on the
assumption that the reaction observed is $\piZeroRxn$.

\item Identify the $\pi^0$ meson missing mass peak.

\item Determine the yield for the $\pi^0$ meson by subtracting any
background from beneath the meson peak.

\item Correct the meson yield for spectrometer acceptance and detection 
efficiency based on Monte Carlo estimates of those quantities.  This
Monte Carlo technique was compared to empirical measurements of the
CLAS acceptance and detection efficiency for a single proton, using
measurements of the reaction $\ppipiRxn$ made concurrently with the
data discussed here, to validate the Monte Carlo simulation used.

\item Normalize the yield using a measured absolute photon
flux normalization procedure.

\end{list}

In the following sections, each of these steps is described.  Also
presented are sample results, and, in some cases, tests that establish
the validity of the procedures used.

\subsection{Recoil proton and pion identification; kinematic variables}
\label{sec:Event}


\begin{figure}
\includegraphics[scale=0.45]{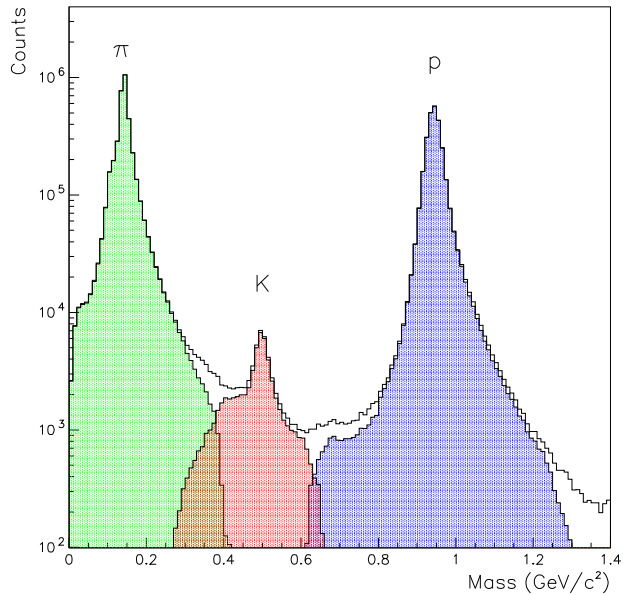}
\caption{Mass distribution of the detected charged hadrons
(note logarithmic scale).
\label{fig:pid}}
\end{figure}
 
The tracking information provided by the drift chambers within CLAS
gave momentum and scattering angle information on charged particles
scattered within the detector volume.  Time-of-flight and start
counter information, coupled with the track information provided by
the drift chambers, yielded velocity and momentum determination.

Particle identification in this analysis was performed using the GPID
algorithm \cite{gpid}.  The method uses the momentum of the detected
particle, and sequentially calculates trial values of $\beta$ for the
particle for possible particle identities.  Each one of the possible
identities is tested by comparing the trial value of $\beta$ for a
given particle type to the empirically measured value of $\beta$ (as
determined by CLAS tracking and time-of-flight information).  The
particle is assigned the identity that provides the closest trial
value of $\beta$ to the empirically measured value of $\beta$.
Figure\ \ref{fig:pid} shows the mass distribution of the identified
positively charged particles.  The GPID algorithm also attempts to
find a matching photon in the tagging system for every charged
particle detected in CLAS.  A matched photon means that there was one
and only one tagged photon in the trigger window, which, in this
analysis, is defined as being within the trigger coincidence window.
Particles that are determined not to have a matching photon are
considered to be a measure of the accidentals (to be described in more
detail in the next subsection).

Geometrical fiducial cuts in each of the six sectors of CLAS were
imposed on all protons.  The region selected for accepting protons in
each sector corresponded to a region of relatively uniform detection
efficiency versus azimuthal angle.

\subsection{Missing mass reconstruction} 

\begin{figure}
\includegraphics[scale=0.45]{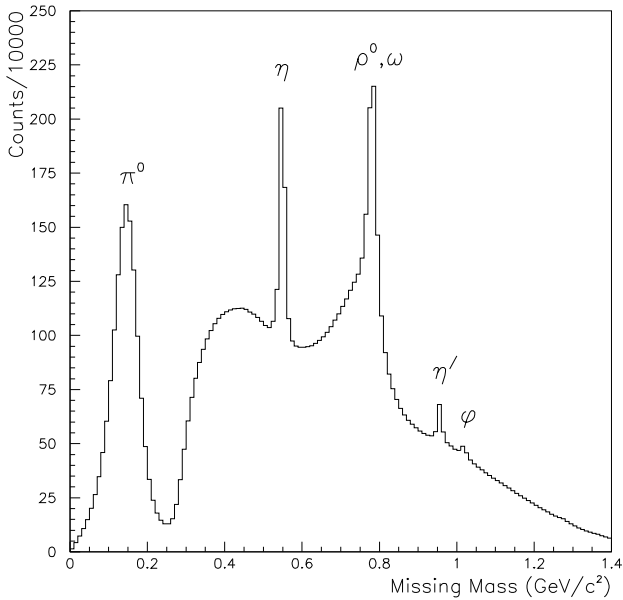}
\caption{ Missing mass spectra obtained using CLAS,
assuming the reaction $\xRxn$.
\label{fig:mm}}
\end{figure}

The momentum determined by CLAS was corrected for energy loss within
the cryogenic target cell to reconstruct the momentum of the detected
proton at the reaction vertex within the cryogenic target.  In
addition to the energy loss correction, a further momentum and photon
energy correction developed by Williams \textit{et al.}  \cite{momcor}
for $g1c$ data was applied.  (This second momentum correction was
required to correct for problems with the magnetic field map
associated with the CLAS detector.)  The measured scattering angle and
momentum can be used to construct a missing mass based on the
assumption that the reaction observed is $\xRxn$, where $X$ is the
undetected particle in the two-body final state.  Based on this
assumption, the missing mass spectrum of the full spectrometer
acceptance for all photon energies is shown in Fig.\ \ref{fig:mm}.
The $\eta$, $\rho^0$/$\omega$, and $\eta^{\prime}$ peaks are clearly
seen atop a background dominated by multi-pion events.  The $\pi^0$
peak is clearly discernible.

Taking each proton event that did not have a matching incident photon
as noted above, and integrating over all of the out-of-time (not
within the trigger coincidence window) incident photons for that
event, determined the distribution of accidental coincidences between
the CLAS and the photon tagger, under the assumption that coupling the
out-of-time tagger events for each non-matched proton created a fair
representation of the accidental coincidences between the CLAS and
tagger.

A failure to match a particle to an incident photon mainly occurs when
reconstructed timing information for the track or tagging system is
missing.

\subsection{Distribution of events into kinematic bins}

The events, from the 2.445 and 3.115 GeV data sets, constituting the
full missing mass spectrum described in the previous section were
distributed into bins in photon energy $\Eg$ and $\cosThetaCmPi$.  The
widths of these ``kinematic bins" ($\Delta\Eg = 50$ MeV in photon
energy and $\Delta\cosThetaCmPi$ = 0.1) were chosen such that, in
general, there were at least 1000 $\pi^0$ events in each kinematic
bin.

\subsection{Determination of pion yield within each kinematic bin}

\begin{figure}
\includegraphics[scale=0.45]{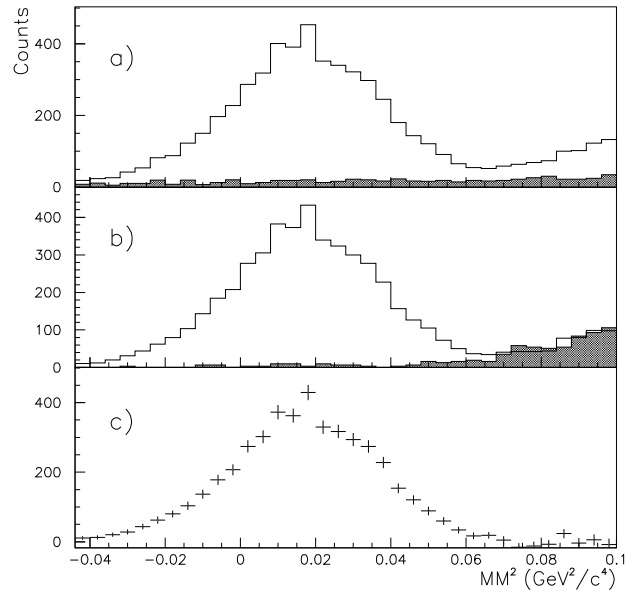}
\caption{ $\pi^0$ meson yield extraction for the kinematic bin with
$\Eg = 1.425$ GeV and $\cosThetaCmPi$ = 0.45.  The top panel a) is the
missing mass yield for this bin, with the accidental contribution
displayed as a shaded region.  The middle panel b) shows the missing
mass distribution with the accidentals subtracted, and the shaded
region represents the $2\pi$ contribution.  The bottom panel c) shows
the extracted $\pi^0$ yield after both contributions have been
subtracted from the missing mass distribution.
\label{fig:fit1400_6}}
\end{figure}

The $\pi^0$ yield within each kinematic bin was determined by
subtracting the background under the peak in the missing mass
spectrum.  We proceeded with the assumption that the background in the
missing mass spectra arose from two particular sources:
\newcounter{bean2}
\begin{list}{\arabic{bean2}.}
{\usecounter{bean2}
\setlength{\rightmargin}{\leftmargin}}
\item Accidental coincidences between CLAS and the photon tagger.
\item Two pion photoproduction via the reaction $\xRxn$, where $X =
\pi^+ \pi^-$.
\end{list}
The accidental contributions to each kinematic bin were determined as
described in Section III.B.  Since events with $p\pi^{+}\pi^{-}$ final
states were copiously produced in this experiment (as may be seen in
Fig.\ \ref{fig:mm}), the shape of the low energy portion of the $2\pi$
background contribution could be reliably determined.  This shape was
used to generate the background beneath the $\pi^0$ peak, which was
then subtracted from the $\pi^0$ yield for each kinematic bin.  An
example of the individual contributions to the background beneath the
pion peak can be seen in Fig. \ref{fig:fit1400_6}.

\subsection{Acceptance and efficiency}
\label{sec:Acceptance}


The CLAS detector acceptance and detection efficiency for recoil
protons were measured and then compared to a Monte Carlo simulation.
The reaction used to determine the empirical acceptance and efficiency
of protons in CLAS was $\ppipiRxn$. 
This reaction was used due to the high number of events for that final
state and, since all the final products leave charged tracks in the
CLAS, these events are easily observed.
Both pions were required to be detected in the event and
both matched to the same photon.

\begin{figure}
\includegraphics[scale=0.45]{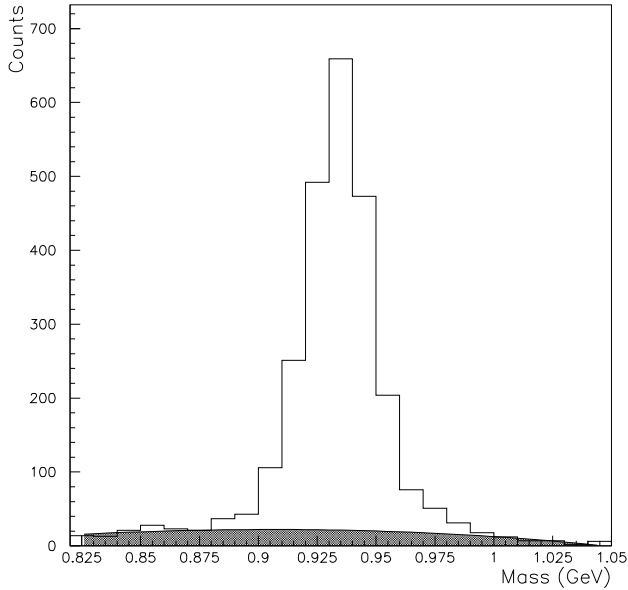}
\caption{ Missing mass (mass of $X$) for the reaction 
$\gamma \ p \rightarrow \ \pi^- \ \pi^+ X$ near the mass of the proton
(shaded region represents the background).
\label{fig:acc1550_7}}
\end{figure}

The data used for the empirical acceptance and efficiency calculations
included only events where two and only two charged pions were
detected in the CLAS.  For each event, a missing mass reconstruction
from the kinematical information from the two pions was performed to
determine if a proton should have been seen in the CLAS.  As shown in
Fig.\ \ref{fig:acc1550_7}, the proton generally was very cleanly
defined, so that the determination that a proton should have been seen
could be made without ambiguity.  The background beneath this peak was
approximated by a third order polynomial and subtracted.  The same
fiducial cuts applied to the protons noted above were applied to both
reconstructed and CLAS-identified protons.

The proton acceptance and efficiency $\alpha$ empirically determined
in this fashion is the ratio of the number of protons ``seen'' $N_{s}$
(i.e., identified in CLAS through the normal particle identification
procedure, GPID, in the empirical data set) and the total number of
events ${N_{p}}$ where a proton should have been seen, based on
reconstructed four-vectors from the pions, detected in the same subset
of events.

A comparison of Monte Carlo events to actual data for the $\ppipiRxn$
reaction (rebinned for the $\piZeroRxn$ reaction) was performed.
Simulated events were obtained by generating $10^7$ $\ppipiRxn$ events
that were isotropic in phase space and then processed through a full
GEANT simulation of CLAS.  These events were then ``smeared'' to
simulate the drift chamber and time-of-flight resolution.  After
smearing, the events were processed as normal data.

\begin{figure}
\includegraphics[scale=0.45]{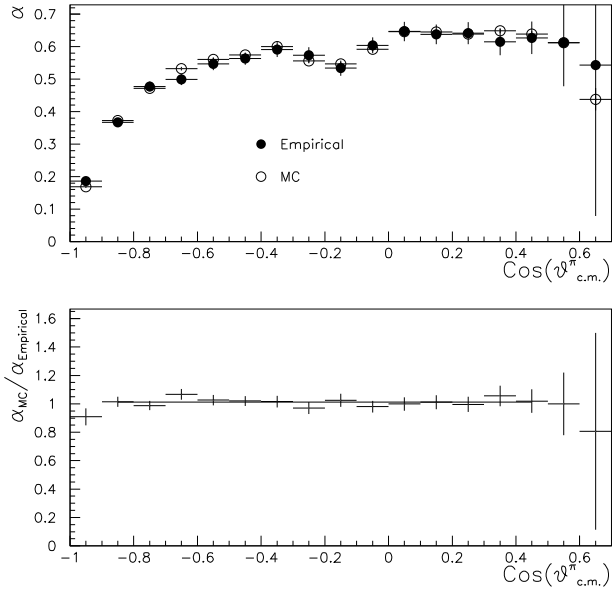}
\caption{ Empirical and Monte Carlo acceptance and efficiencies $\alpha$ for 
$\pi^0$ photoproduction for $\Eg = 725$ MeV.  The top panel shows
$\alpha$ (filled circles represent the empirical method and open
circles represent the Monte Carlo method).  The bottom panel shows the
ratio of $\alpha$ (Monte Carlo method divided by the empirical
method).
\label{fig:effComp}}
\end{figure}
 
A comparison between the empirical and Monte Carlo data can be seen
for an incident photon energy of 725 MeV in Fig. \ref{fig:effComp}.
To perform a statistical comparison for these two data sets, the
bottom panel of Fig.\ \ref{fig:effComp} (Monte Carlo/empirical) was
fit with a polynomial of zero order over the $\cosThetaCmPi$ range of
-0.85 to 0.45. For angles above this range, there were not enough
empirical events, after rebinning to the $\pi^0$ reaction, to perform
a reliable comparison.  For $\cosThetaCmPi < -0.9$, the Monte Carlo
did not agree well with the empirical method, and, for this reason,
points with $\cosThetaCmPi < -0.9$ were left out of the subsequent
steps of analysis.  The zero-order result of the fit is called the
``acceptance and efficiency ratio.''  The ratio of empirical to Monte
Carlo for each energy and $\cosThetaCmPi$ bin (within 0.65 GeV $\le
E_{\gamma} \le$ 1.8 GeV and $-0.9 < \cosThetaCmPi < 0.5$) was placed
in a histogram, and then fit with a Gaussian.  The center of the
Gaussian was found to be $1.0012 \pm 0.002$, with standard deviation
$0.0312 \pm 0.016$, and reduced $\chi^2 = 0.76$, demonstrating our
ability to find and correct for inefficiencies and acceptance.

With this confirmation of the Monte Carlo validity, for $-0.9 <
\cosThetaCmPi < 0.5$ and 0.65 GeV $\le E_{\gamma} \le$ 1.8 GeV (and
assuming the validity for all other energies and $-0.9 < \cosThetaCmPi
< 0.9$), the acceptance and efficiency for each kinematic bin for the
reaction $\piZeroRxn$ was obtained by generating $10^7$ events
(weighted by the cross sections given by the SAID
solution~\cite{sm02}).  The ratio of processed events, to the number
generated, for a given bin, served as a measure of the acceptance and
efficiency.

\begin{figure}
\includegraphics[scale=0.45]{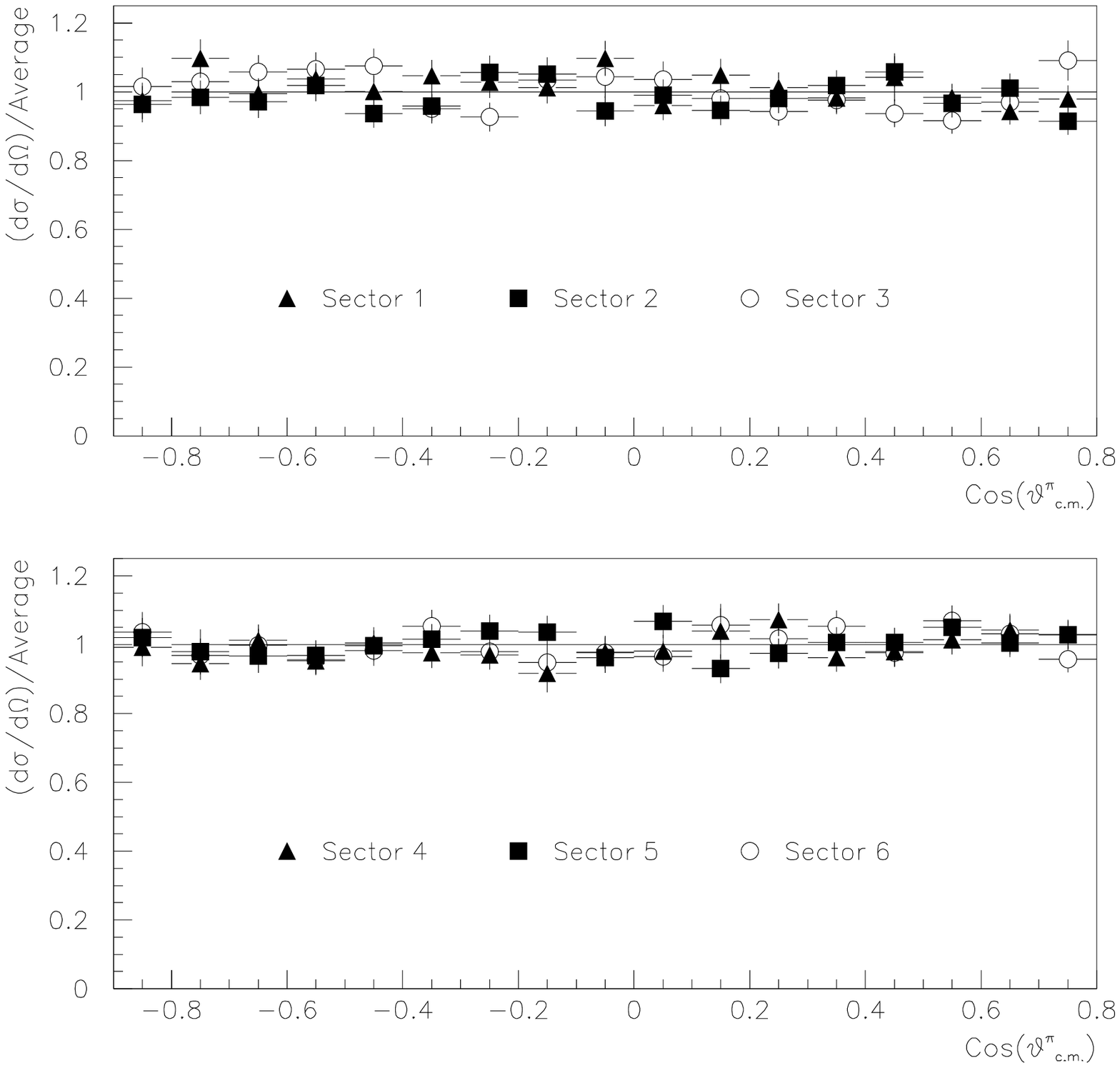}
\caption{Differential cross sections for each sector divided by the
average over all CLAS sectors ($\Eg = 1.025$ GeV) for the reaction
$\piZeroRxn$.
\label{fig:sectorDcs1000}}
\end{figure}

\subsection{Sector-by-sector comparison}
The CLAS has six sectors that are nominally identical, but these
sectors, in fact, differ relative to each other in acceptance and
detection efficiency due to differences that either initially existed
in their construction or by various changes and differences that have
arisen over time since they were first installed.  A sector-by-sector
comparison of the differential cross sections was performed to check
the consistency of the extracted cross sections.  Since the Monte
Carlo simulation should reflect sector-by-sector changes in the
detector arising from, for example, broken drift chamber wires and
time-of-flight paddles, a sector-by-sector comparison of the
differential cross section explores the reliability of the Monte Carlo
simulation with respect to these detector irregularities.  A typical
sector comparison plot is shown in Fig.\
\ref{fig:sectorDcs1000}.  The plot shows the differential cross
sections of each sector of CLAS divided by the average over all
sectors.  An examination of these plots for each energy revealed no
systematic shifts visible in the differential cross section for either
$\Eg$ or $\cosThetaCmPi$.

\subsection{Bin migration}

To estimate the systematic error associated with bin migration, the
acceptance and efficiency results calculated using SAID-weighted
events were compared to acceptance and efficiency results using
non-weighted events.  For each kinematic bin, the systematic error
associated with bin migration would be the uncertainty in the
corrected value multiplied by the amount of the correction.  Since the
amount of the correction was found to be typically less than 1.0\%,
and always less than 2\%, the systematic uncertainty associated with
bin migration was assumed to be ignorable.

\subsection{Trigger inefficiency}

The determination of a charged particle trigger inefficiency for the
data was performed by looking at data from a running period that had a
CLAS-detected photon event trigger condition in addition to the
CLAS-detected charged particle event trigger described above for the
g1c running period.  An experimental running period that had both
charged particle and photon triggers occurred just before the running
period used in this analysis. (This dual trigger running period was
called g2a; a more detailed discussion of the g2a experiment can be
found in Ref. \cite{g2a}.)  By observing events in the g2a running
period that had a photon trigger and no charged trigger, yet clearly
had a proton detected by CLAS in the event, the inefficiency of the
charged particle trigger in CLAS for protons could be determined.
This correction was applied to each kinematic bin and was always less
than 1.0\%.

\subsection{Normalization}
\label{sec:Norm}


In contrast to many of the previously published data sets for
$\piZeroRxn$, the data in this analysis were not normalized to
previous data for the process or to the SAID solution for any reaction
channel.  The {\em{absolute photon flux}} for the entire tagger photon
energy range was determined by measuring the rate of scattered
electrons detected in each counter of the focal plane of the
bremsstrahlung photon tagger by sampling focal plane hits not in
coincidence with CLAS.  The detection rate for the scattered electrons
was integrated over the lifetime of the experiment and converted to
the total number of photons on target for each counter of the tagger
focal plane.  The tagging efficiency was measured in dedicated runs
with a total absorption counter (TAC) downstream of the cryogenic
target, which directly counted all photons in the beam. The details of
the method can be found in Ref. \cite{gflux}.

\section{Uncertainties}
\label{sec:Errs}


Having investigated various sources of uncertainties in the analysis,
we collect and summarize here the various uncertainties determined for
the cross sections obtained in this work.

\subsection{Trigger inefficiency}
From the \textit{trigger inefficiency} study given in subsection
III.H, an overall estimated systematic uncertainty of 1\% for the
trigger inefficiency correction factors was taken as a very
conservative estimate of the systematic uncertainty associated with
the trigger inefficiency.

\subsection{Background subtraction}
The uncertainty associated with the \textit{background subtraction} is
purely statistical, and these were taken into account on a bin-by-bin
basis.  No systematic uncertainties for the background subtraction are
included.

\subsection{Effects of momentum correction}
As noted in Section III.B, a pair of momentum corrections for the
rescattered proton was made prior to forming the missing mass spectra.
These momentum corrections affect extraction of the $\pi^0$ from the
missing mass distribution, and can alter the the center-of-mass angle
for the scattered proton, sometimes transferring events from one
kinematic bin to another (``bin migration").  An estimate of the
systematic uncertainty introduced by effects of these momentum
corrections can be formed by looking at the magnitude of the momentum
correction and the magnitude of the resulting deviation from the
optimal value determined by a pull distribution.

To determine the uncertainty associated with the momentum correction,
the reaction $\ppipiRxn$ was studied using methods described in
greater detail in Ref. \cite{momcor}.  The variable $z$ is defined as
$z = \Delta p_p / \sigma = \left( p_{p(fit)} - p_{p(measured)}
\right)/\sigma$, where $p_{p(fit)}$ represents the best value of the
momentum as determined by a kinematic fit for data which does not
include the CLAS measured value for the proton momentum, and $\sigma$
is the standard deviation. The distribution of $z$ was examined by
histograming and fitting the results to a Gaussian. Histograms were
generated for $z$ from events with and without the momentum correction
for the proton.  The difference between these means is taken as a
measure of the average momentum correction.

The initial (final) mean and standard deviation of the pull
distribution were 0.208 and 0.963 (-0.022 and 1.011).  The magnitude
of the change in the mean before and after the momentum correction is
0.23.  The amount that the mean of the distribution is still different
from the optimal value of zero is 0.022. This suggests that the
momentum correction is only good up to 0.022/0.23 = 0.096. Thus, we
estimate conservatively that the systematic uncertainty in the
momentum correction is 10\%.

The resulting uncertainty in the cross section due to the uncertainty
in the \textit{momentum correction} is taken to be 10\% of the
absolute value of the difference between the momentum corrected and
non-momentum corrected differential cross sections.  This uncertainty
is typically less than a few percent and is added in quadrature on a
bin-by-bin basis.

\subsection{Photon flux normalization}

Due to the large number of out-of-time photons used to obtain the
photon flux normalization \cite{gflux}, the statistical uncertainties
associated with the \textit{photon flux normalization} are always far
below 1\%.  For this reason, an overall 1\% uncertainty for the
statistical error of the normalization is included as a very
conservative estimate of this uncertainty.

Since data collection intervals were taken with two different incident
electron beam energies, it is reasonable to compare whether any
systematic differences arise between the sets of data taken at these
two energies.  To make this comparison, a histogram was generated for
the ratio $R$, obtained by taking the differential cross sections of
the 3.115 GeV data set and dividing those values by the cross sections
obtained in the 2.445 GeV data set, for each kinematic bin. A Gaussian
was fit to this histogram, with a resulting centroid of 0.996 $\pm$
0.002.  From this comparison, we estimate the systematic uncertainty
introduced by combining data sets from the two electron energies to be
0.4\%.  A fluctuation in the photon flux for differing endpoint
energies could be due to the energy calibration of the tagger
(including effects due to possible, slight differences in the tagger
magnet field map).  Since an increase of 27\% in initial electron
energy (2.445 GeV to 3.115 GeV) causes only a 0.4\% standard deviation
in the photon flux, a conservative estimate of 1\% for the systematic
uncertainty in photon flux associated with these
\textit{``field-to-field''} differences is used.

The largest source of uncertainty in the normalization is caused by
the ``tagger efficiency'' \cite{tag}. The tagger efficiency is
essentially a measure of the amount of the tagged photon beam that
survives collimation, as determined during normalization runs.  The
value of the tagger efficiency is dependent upon the electron beam
supplied by the accelerator, and will vary on a run-by-run basis
determined by the condition of the electron beam tune.

To estimate the effects of fluctuations in the electron beam incident
on the radiator of the tagger, we calculate the proton yield,
normalized to incident photon flux, for each run, and determine the
mean and standard deviation of this normalized yield (assuming
Gaussian statistics). The \textit{run-by-run uncertainty} was
estimated (see Table \ref{tbl:systematic} ) using the standard
deviation of this normalized proton yield, dividing by the mean, to
get the fractional variation.

\subsection{Absolute normalization}
The systematic uncertainty for the
\textit{absolute normalization} is comprised of five parts:

\newcounter{bean4}
\begin{list}{\arabic{bean4}.}{\usecounter{bean4}\setlength{\rightmargin}{\leftmargin}}
\item uncertainty in the cryogenic target density  
(discussed in detail in Ref.~\cite{targetDensity}) and length; 
\item statistical error of the normalization (described in the
preceding subsection);
\item run-to-run variations in the normalized proton yield (described in the
preceding subsection);
\item uncertainty associated with the tagger magnetic field calibration 
(``field-to-field" differences) in combining the results for the two
incident electron energies (described in the
preceding subsection); and the
\item uncertainty of the trigger inefficiency (described in subsection III.H).
\end{list}

Table \ref{tbl:systematic} shows the values for these contributions to
the systematic uncertainty of the absolute normalization. After adding
items (1), (2), and (3) in quadrature and then linearly adding the
remaining contributions, the systematic uncertainty for the
normalization is 3.7\% and 3.9\% for the 2.445 GeV and the 3.115~GeV
data sets, respectively.  Even when all of the systematic
uncertainties of the normalization are simply added together, the
resulting uncertainty is 4.4\% for the 2.445 GeV data set, and 4.8\%
for the 3.115~GeV data set.  Since the estimated error is never over
5\%, the overall systematic uncertainty in the absolute normalization
is estimated as having a conservative upper bound value of 5\%.

\begin{table}[ht]
\caption{Systematic uncertainties in the absolute
normalization.}\label{tbl:systematic}
\vspace{2mm}
\centering
\begin{tabular}{|c|c|c|c|}
\hline
Data Set &  Run-to-Run   &  Target density  &  Target length    \\
\hline
2.445 GeV & 0.98\% & 0.14\% & 0.28\% \\
3.115 GeV & 1.25\% & 0.28\% & 0.28\%  \\
\hline
Data Set &  Field-to-field   &  Trigger  & Statistical    \\
\hline
2.445 GeV & \multirow{2}{*}{$<1$\%} & $<1$\% & $<1$\% \\
3.115 GeV & & $<1$\% & $<1$\% \\
\hline
\end{tabular}
\end{table}

\section{Results}
\label{sec:Results}

The differential cross sections obtained here are compared with
previous data from MAMI-B~\cite{Mainz}, GRAAL~\cite{GRAAL}, and
CB-ELSA~\cite{Bonn} in Figs. \ref{fig:g1} through \ref{fig:g3a}.

For incident photon energies E up to 1.275 GeV, the data obtained here
are generally in very good agreement with previous data.
Above that energy discrepancies between this work and the results of the
CB-ELSA Collaboration are seen.
The disagreement begins at forward
meson scattering angles at $E_\gamma$=1.325 GeV, and then begins to
appear at other angles as the energy increases, though the differences
at larger angles are relatively small.  The overall systematic
uncertainty for the CB-ELSA measurements is stated to be 5\% below
1300~MeV and 15\% above that energy.  This compares with the roughly
5\% systematic uncertainty obtained here. With these estimated
uncertainties, the data are in statistical agreement for the larger
angles, while the discrepancies at the smallest angles are larger than
can be accounted for by systematic uncertainties.

Clearly, additional measurements at forward angles are needed to
determine whether the rapid increase suggested by the CB-ELSA data is
correct, or whether the more gradual behavior predicted by the fits
properly describes the trend of the cross section at forward angles.

\section{Amplitude Analysis of Data}
\label{sec:fit}


Multipole amplitude analyses provide a powerful tool for extracting
information about the reaction process in as nearly a
model-independent manner as possible~\cite{sm02}. This approach, in
turn, facilitates the identification of $s$-channel resonances
involved in the reaction process.

Cross sections from this experiment have been included in, and
excluded from, a number of multipole fits to the full SAID
database~\cite{sm02} in order to gauge their influence and
compatibility with previous measurements. In Table~\ref{tab:tbl1}, the
values of $\chi^2$ resulting from these fits are compared to
predictions from MAID (MAID05~\cite{maid} and MAID03~\cite{maid03})
and an earlier SAID analysis (SM02)~\cite{sm02}.  For the purposes of
this discussion, several combinations of data sets and fits are
presented. A fit called ``FDX6" was determined from the world database
such that it includes recent GRAAL~\cite{GRAAL} and
CB-ELSA~\cite{Bonn} data but does not include the present CLAS
dataset, whereas the ``FD16" fit includes the CLAS data. To emphasize
the effects of the CLAS data reported here and to minimize the
influence of CB-ELSA data, the solution ``FA06" is a fit that the
weight factor for this data was artificially increased by a factor of
3.

The fits FD16 and FA06, despite having different weights for the CLAS
data, are in good agreement with each other; this is not surprising,
as the older SM02 fit also follows the CLAS data.  As seen in
Table~\ref{tab:tbl1}, the description of data by MAID is significantly
poorer than by any other fit included in this analysis.

\begin{table}[th]
\caption{$\chi^2$ comparison of fits to
         3~GeV, fit SM02 (to 2.0~GeV)~\protect\cite{sm02}, and two
         recent Mainz fits, MAID05~\protect\cite{maid} and
         MAID03~\protect\cite{maid03} (to
         1650~MeV~\protect\cite{lot}).  See text for
         details. \label{tab:tbl1}}
\vspace{2mm}
\begin{tabular}{|c|c|c|}
\colrule
Solution & Range~(MeV) & $\chi^2$/Data \\
\colrule
FA06     & 3000        &  55640/25524 \\
FD16     & 3000        &  52196/24008 \\
FDX6     & 3000        &  49010/23250 \\
SM02     & 2000        &  35297/17571 \\
MAID05   & 1650        & 141270/21942 \\
MAID03   & 1650        & 486266/21942 \\
\colrule
\end{tabular}
\end{table}
\begin{figure*}[th]
\includegraphics[height=0.85\textwidth, angle=90]{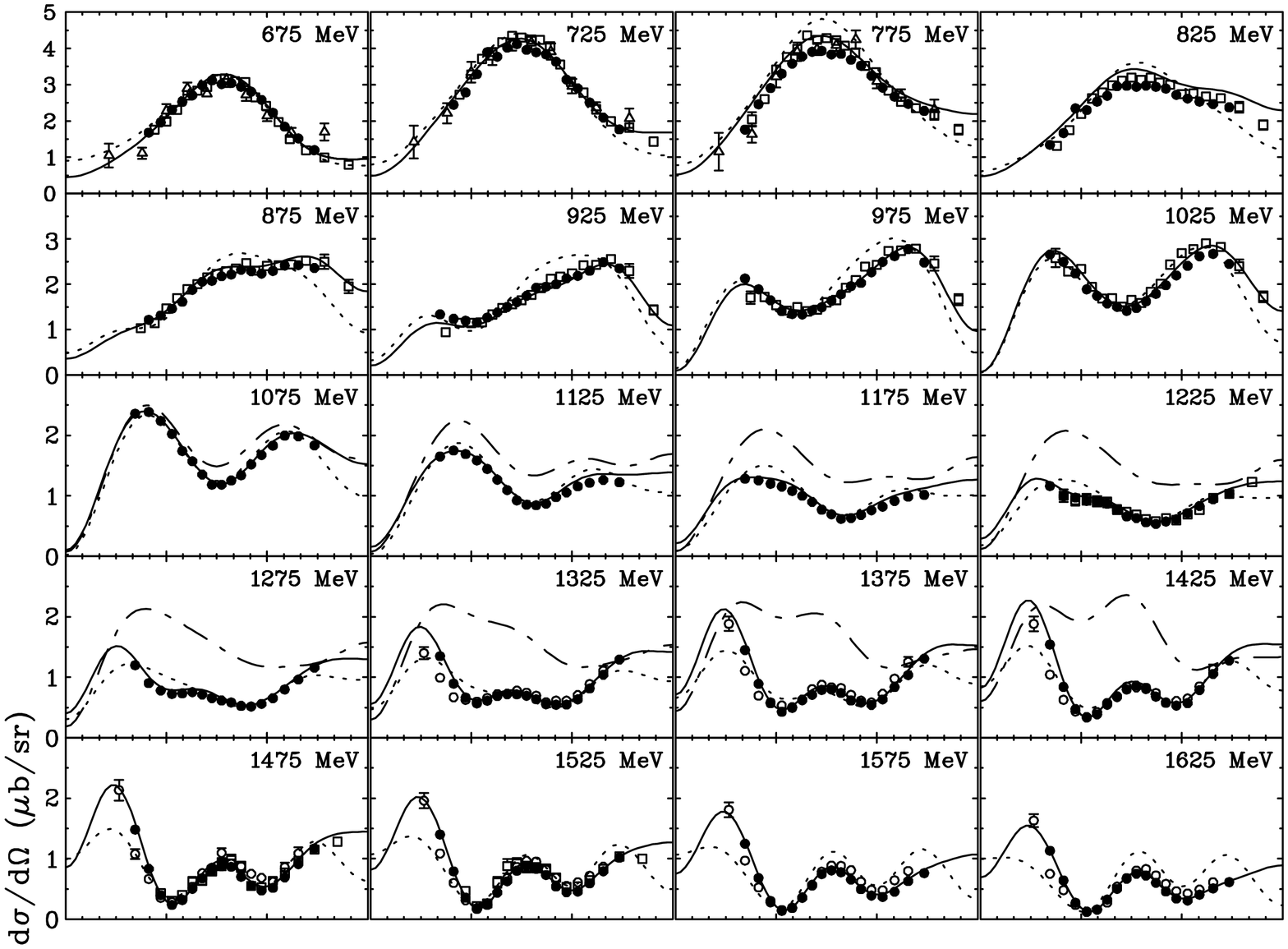}\\
\includegraphics[height=0.85\textwidth, angle=90]{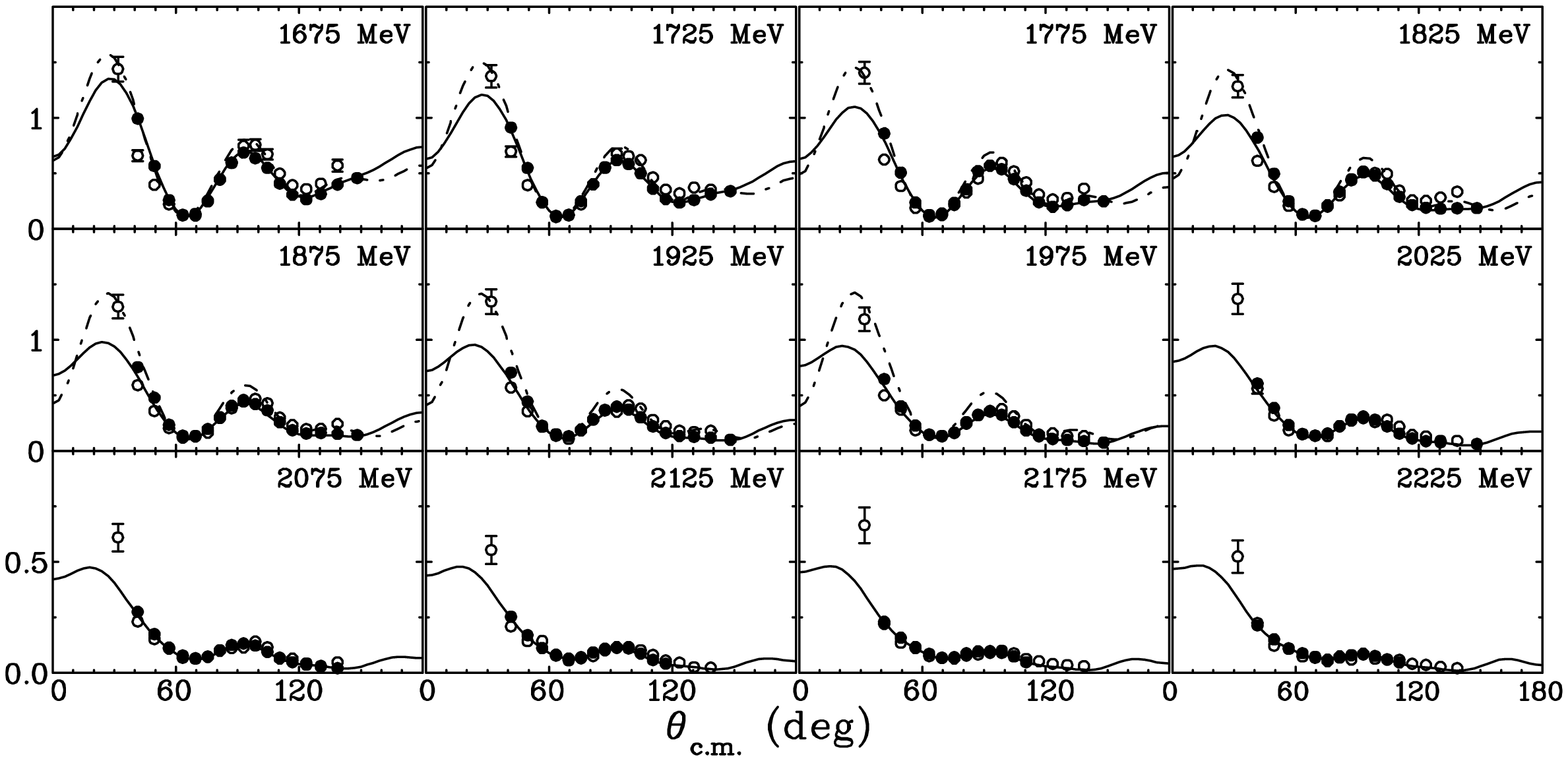}
\caption{The differential cross section for
         $\gamma p\to\pi^0p$ below E$_\gamma$ = 2250~MeV.  The angle
         shown is the pion center-of-mass scattering angle.  Solid
         (short dash-dotted) lines correspond to the SAID FA06
         (SM02~\protect\cite{sm02}) solution.  SM02 curves are shown
         only for E$_\gamma$ between 1650 and 2000~MeV.  Dotted (long
         dash-dotted) lines give the MAID05~\protect\cite{maid}
         (MAID03~\protect\cite{maid03}) predictions.  MAID03 curves
         are shown only for E$_\gamma$ between 1050 to 1450~MeV.
         Experimental data are from the current measurement (filled
         circles), MAMI-B~\protect\cite{Mainz} (open triangles),
         GRAAL~\protect\cite{GRAAL} (open squares), and
         CB-ELSA~\protect\cite{Bonn} (open circles).  The plotted
         experimental data have been selected from energy bins
         spanning at most 5~MeV.  \label{fig:g1}}
\end{figure*}

The FA06 and MAID fits are compared to data over the range of the MAID
analysis in Fig.~\ref{fig:g1}. Above this energy limit (1650~MeV), we
compare FA06 and the older SM02 solution in Fig.~\ref{fig:g1}.  The
deviation of MAID03 from data and the SAID fits, above approximately
1~GeV, is known to the Mainz group. The (preliminary) MAID05 fit is
much improved.  A version to be made publicly available through the
MAID website will contain further modifications~\cite{sabit}.  The
MAID05 and SAID fits are significantly different in the most forward
bump/dip structure and at backward angles.

\begin{figure}[th]
\includegraphics[height=0.40\textwidth, angle=90]{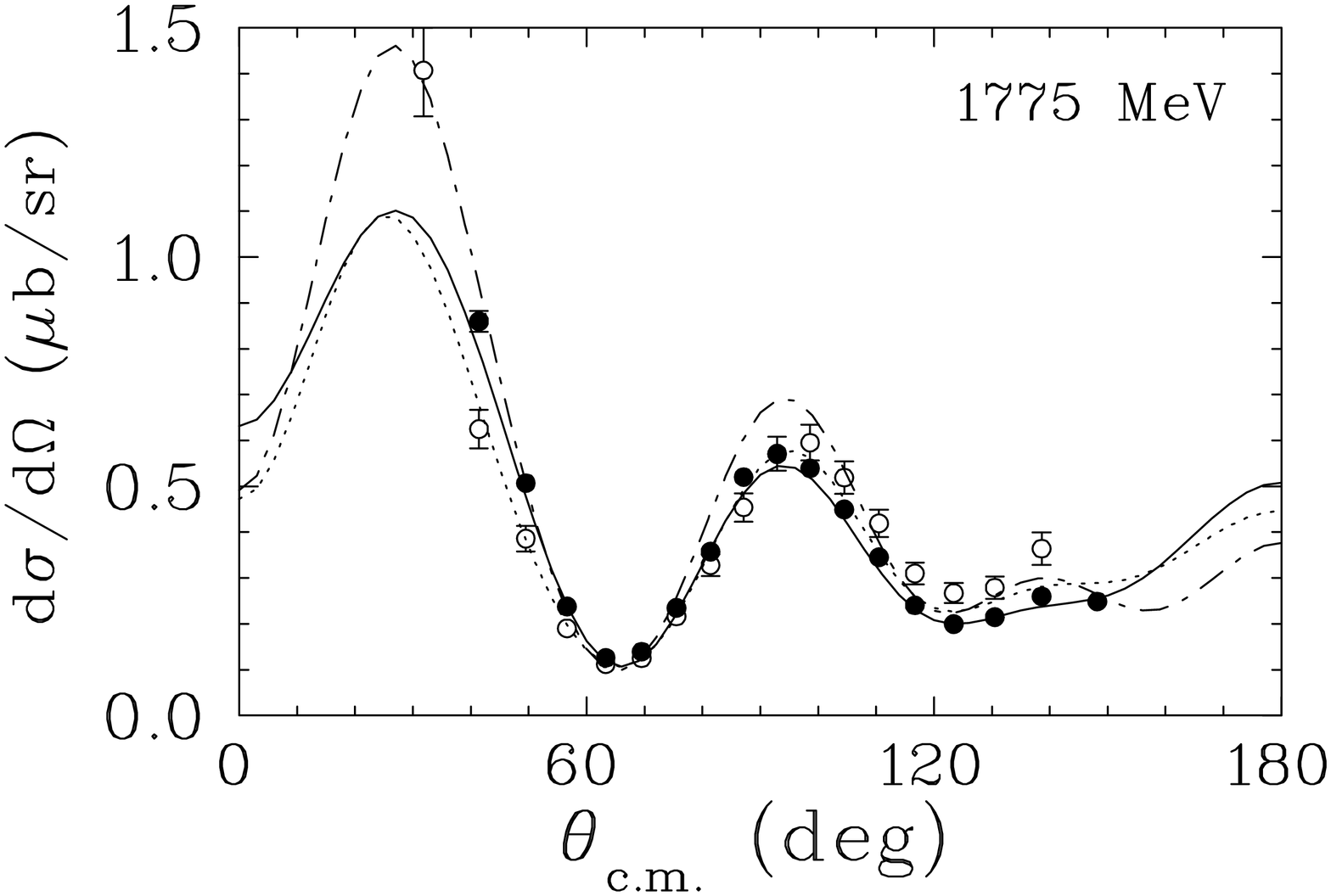}\\
\includegraphics[height=0.40\textwidth, angle=90]{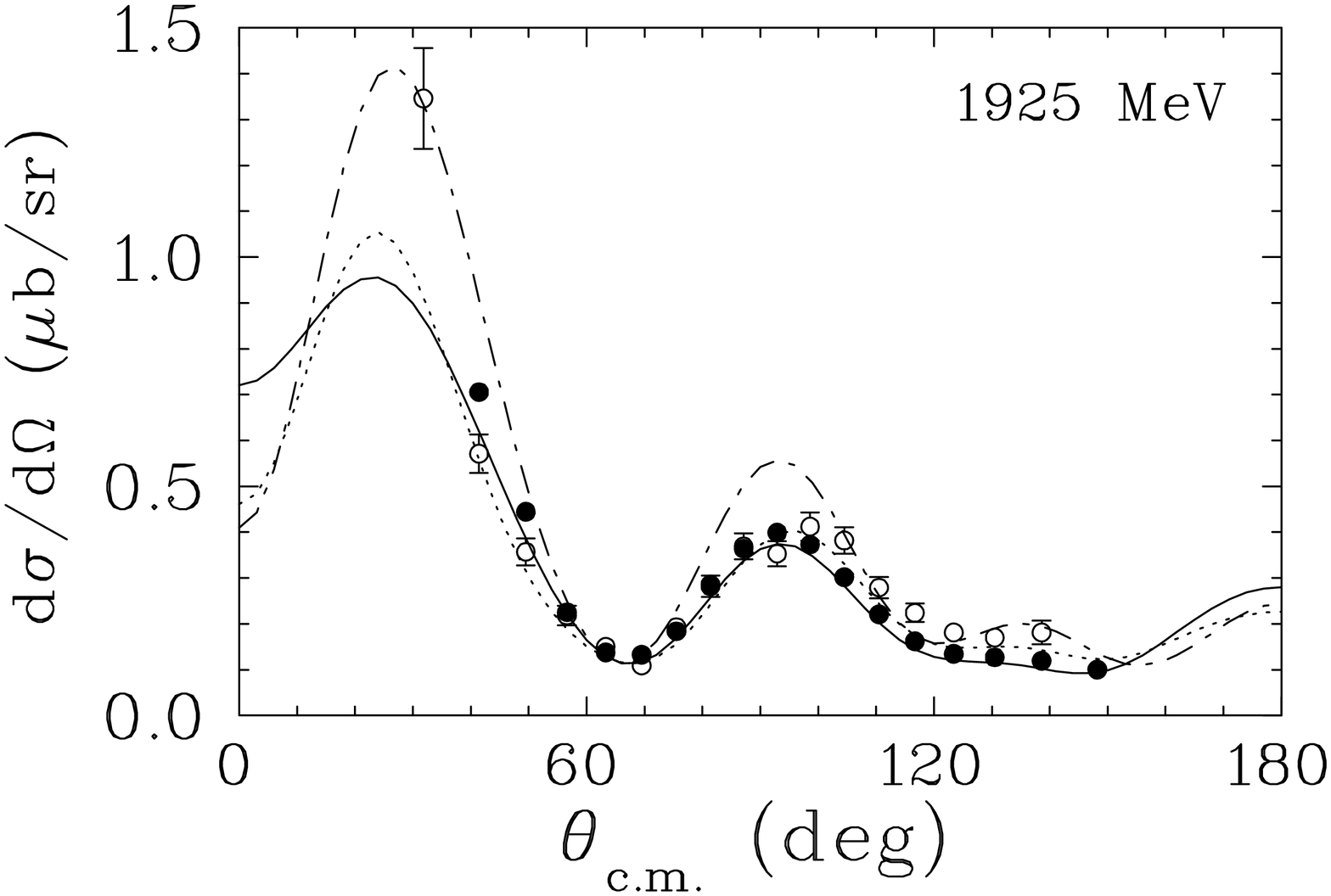}
\caption{The differential cross section for $\gamma
         p\to\pi^0p$ at E$_\gamma$ = 1775 and 1925~MeV. The angle
         shown is the pion center-of-mass scattering angle.  Solid
         (long dash-dotted) lines corresponding to the GW SAID FA06
         (SM02~\protect\cite{sm02}) solution.  Dotted lines represent
         FDX6 results. Experimental data are from the current
         measurement (filled circles) and CB-ELSA~\protect\cite{Bonn}
         (open circles).  \label{fig:g2}}
\end{figure}

The forward region continues to differ most in Fig.~\ref{fig:g2},
where the fits SM02 and FA06 are compared. At intermediate angles,
agreement between the CLAS and CB-ELSA datasets is quite good.  Note
that the older SM02 fit is in perfect agreement with the most forward
CB-ELSA measurements, though these data were not included in the
fit. The model dependence of this forward region is further explored
in Fig.~\ref{fig:g2}.  Note that the FDX6 result, including CB-ELSA
but {\em{not}} CLAS data, is actually in worse agreement with the most
forward CB-ELSA measurements. The FA06 fit (which includes the CB-ELSA
points and the data obtained here) yields results that also fall far
below the most forward point measured by the CB-ELSA
collaboration. While only suggestive, this observation adds some
support to the less rapid increase in the differential cross sections
implied by the data obtained here.

\begin{figure}[th]
\includegraphics[height=0.49\textwidth, angle=90]{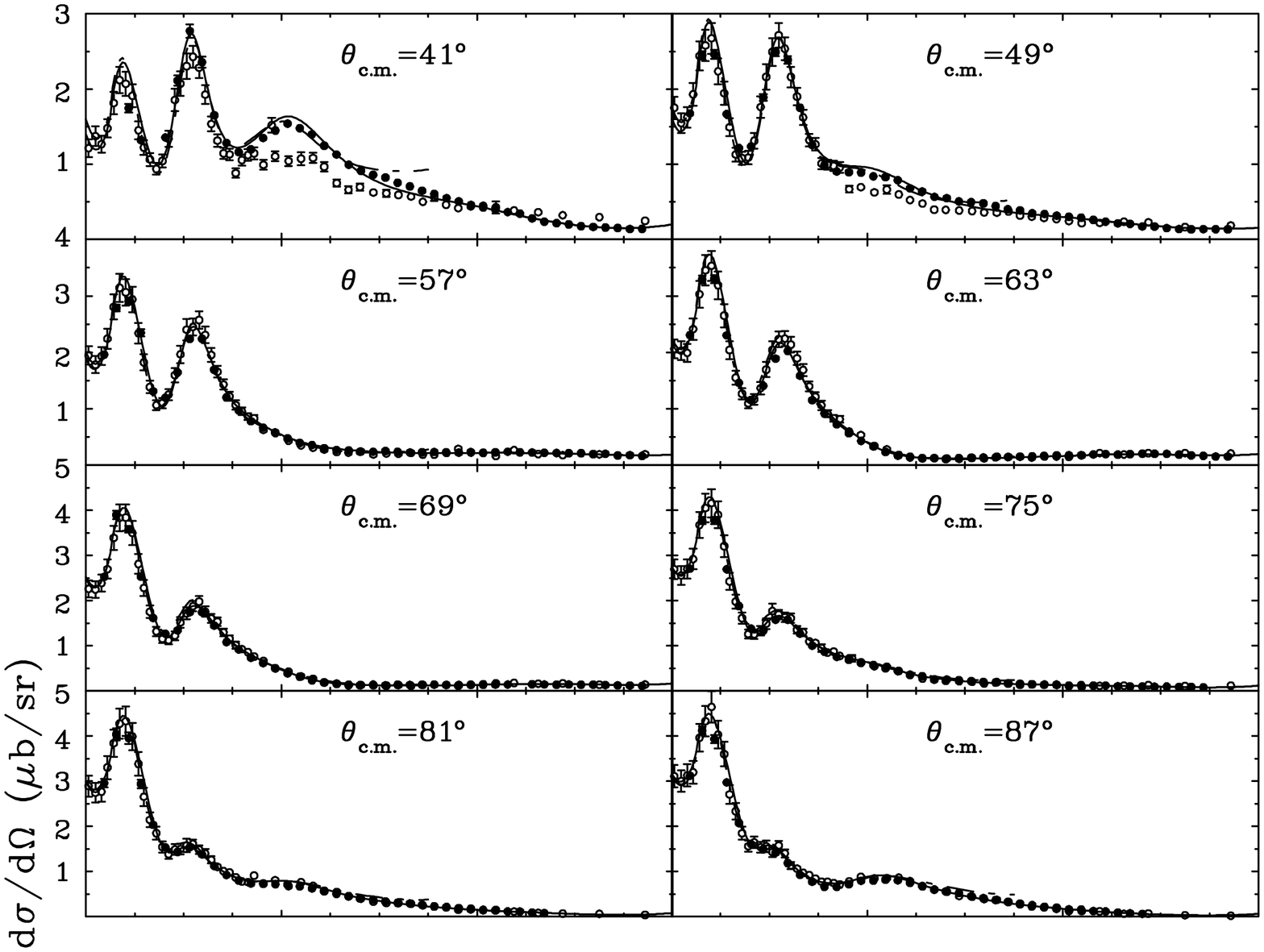}\\
\includegraphics[height=0.49\textwidth, angle=90]{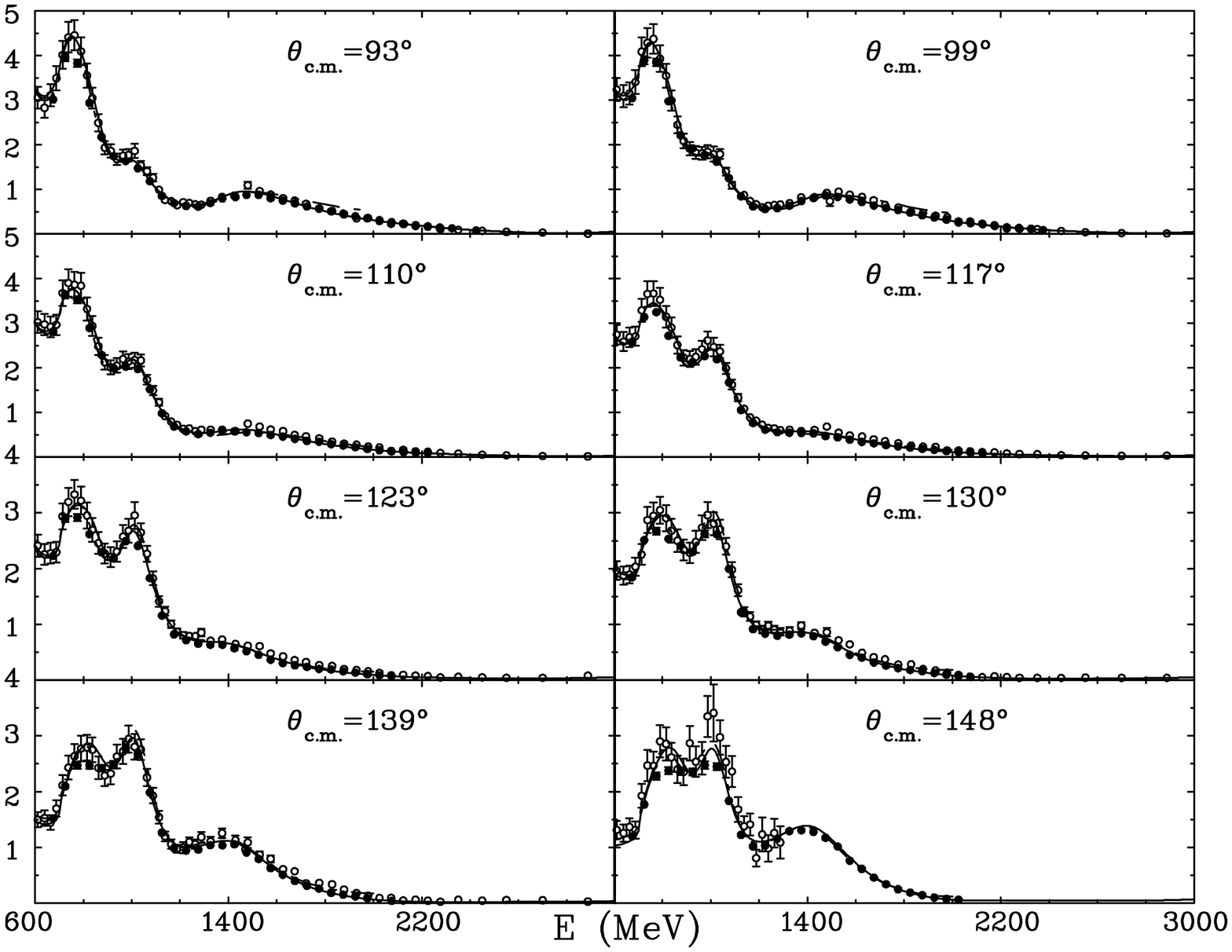}
\caption{Fixed angle excitation functions for
         $\gamma p\to\pi^0p$. The angle shown is the pion
         center-of-mass scattering angle. Solid (long dash-dotted)
         lines corresponding to the GW SAID FA06
         (SM02~\protect\cite{sm02}) solution. Experimental data are
         from the current measurement (filled circles) and
         CB-ELSA~\protect\cite{Bonn} (open circles).  \label{fig:g3}}
\end{figure}

\begin{figure}[th]
\includegraphics[height=0.49\textwidth, angle=90]{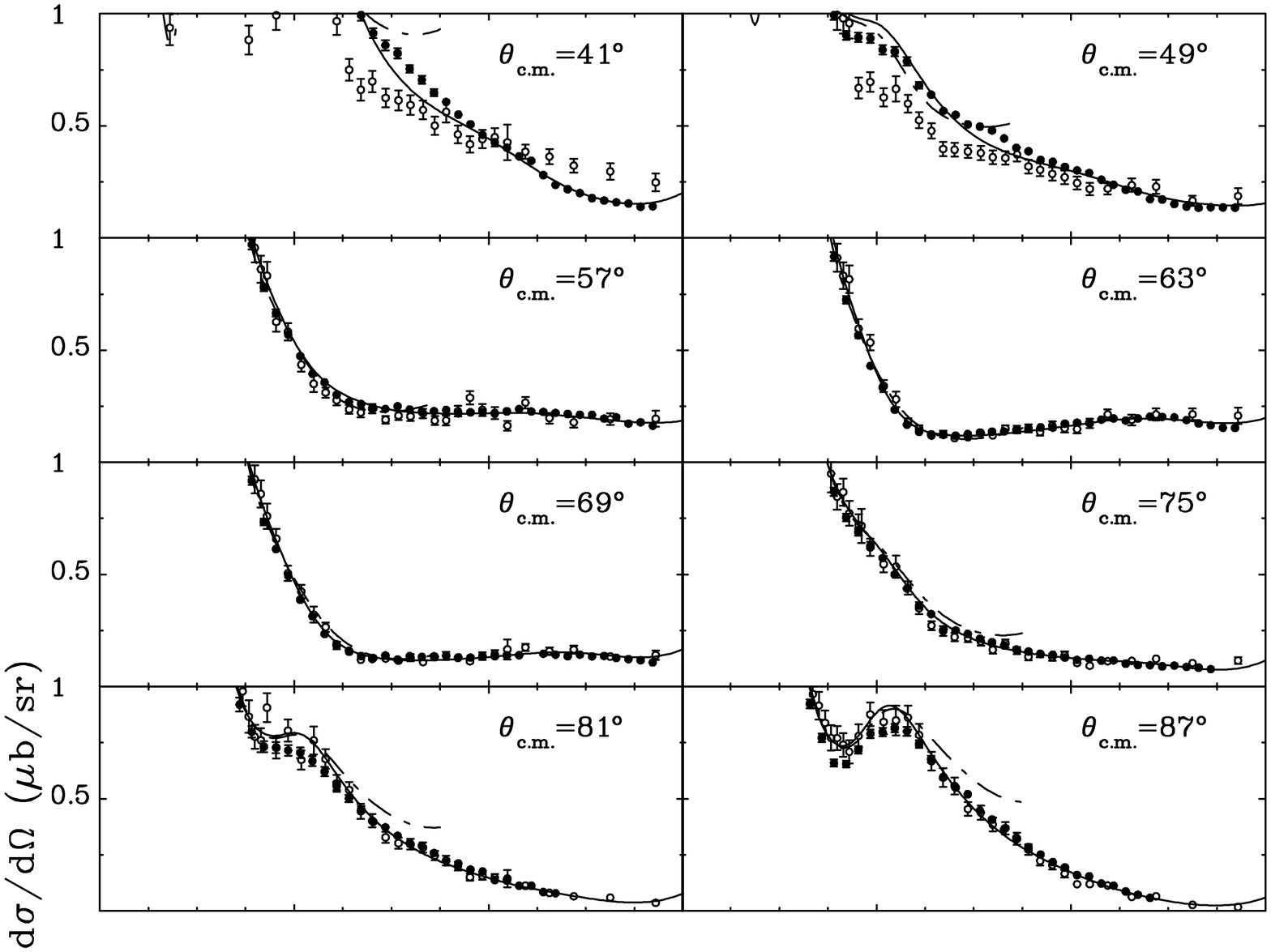}\\
\includegraphics[height=0.49\textwidth, angle=90]{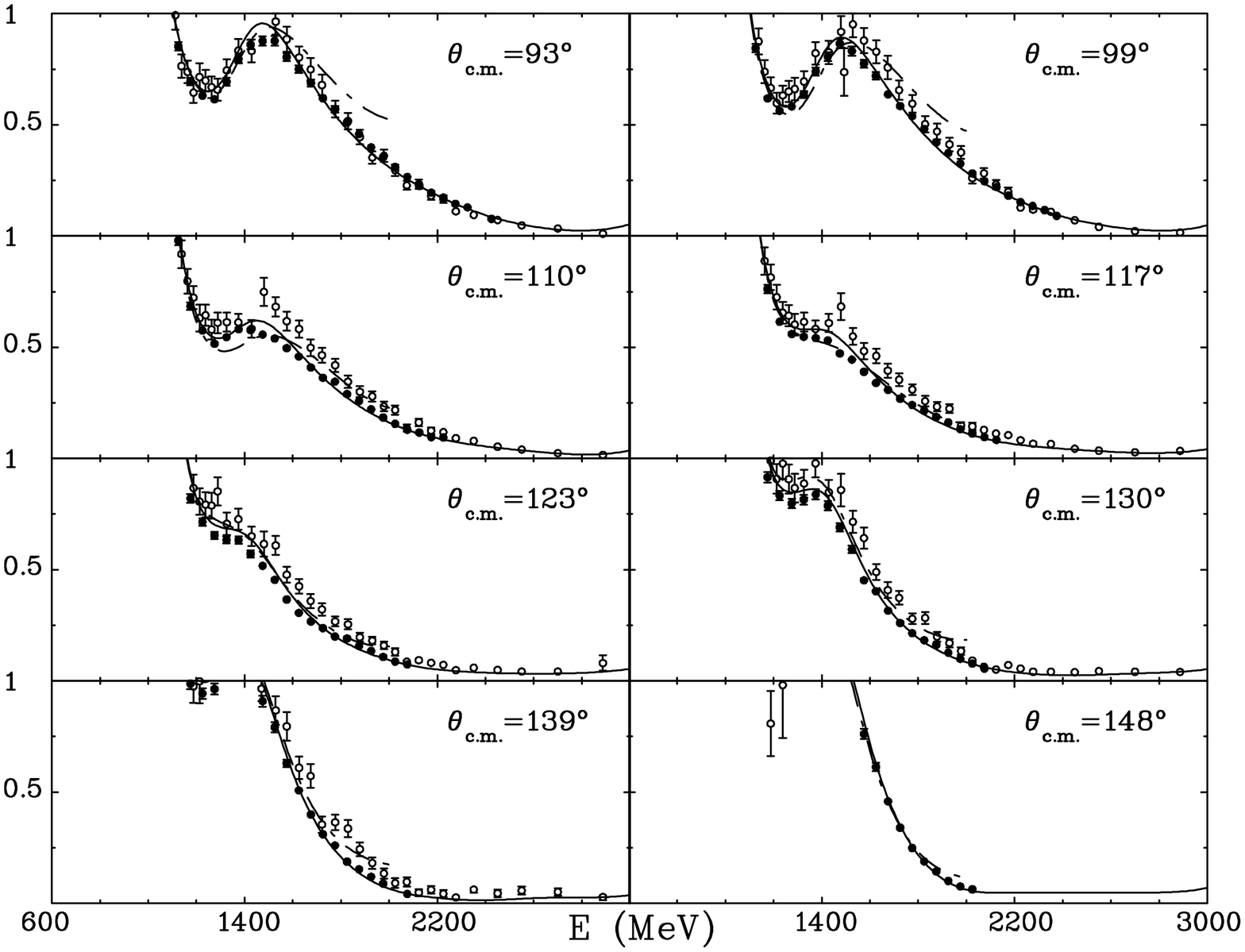}
\caption{Fixed angle excitation functions for
         $\gamma p\to\pi^0p$ below 1$\mu$b/sr. Notation as in
         Fig.~\protect\ref{fig:g3}.  \label{fig:g3a}}
\end{figure}

The present fits have been generated using the most recent GW analysis
of pion-nucleon scattering data, which was extended to cover the full
resonance region~\cite{sp06}. The upper limit on the photon energy is
3~GeV (as compared to 2~GeV for SM02~\cite{sm02}), incorporating the
full CLAS dataset determined here. In Figs.~\ref{fig:g3}
and~\ref{fig:g3a}, we display the energy dependence of cross sections
at fixed angle. This view most clearly shows the kinematic region and
scale of disagreement between the CLAS and CB-ELSA datasets.  Also
apparent is the transition, between 2 and 3~GeV, to a region where the
cross sections have a nearly energy-independent structure.

\section{Resonance Couplings}
\label{sec:ResCoupl}

\begin{figure*}[th]
\centerline{
\includegraphics[height=0.45\textwidth, angle=90]{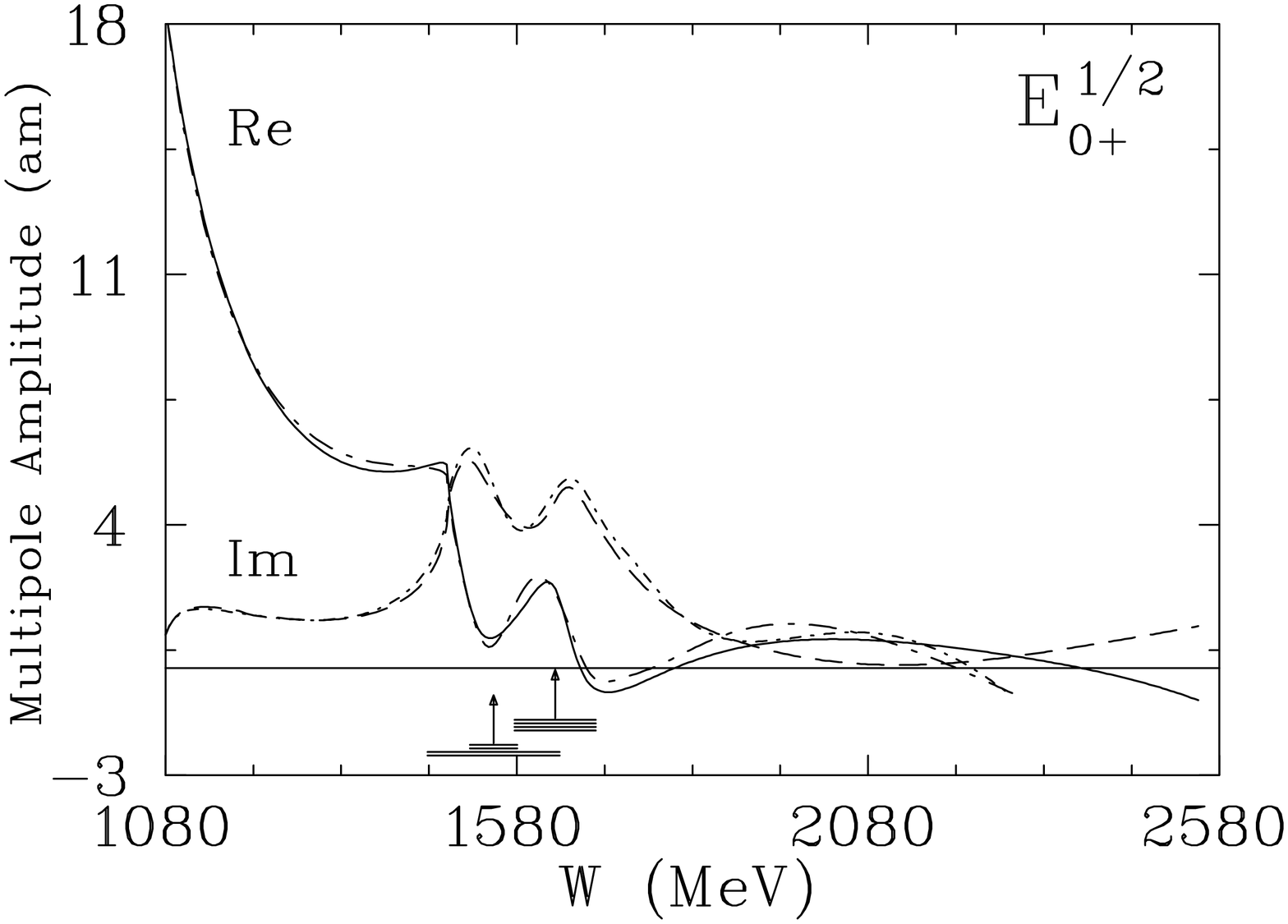}\hfill
\includegraphics[height=0.45\textwidth, angle=90]{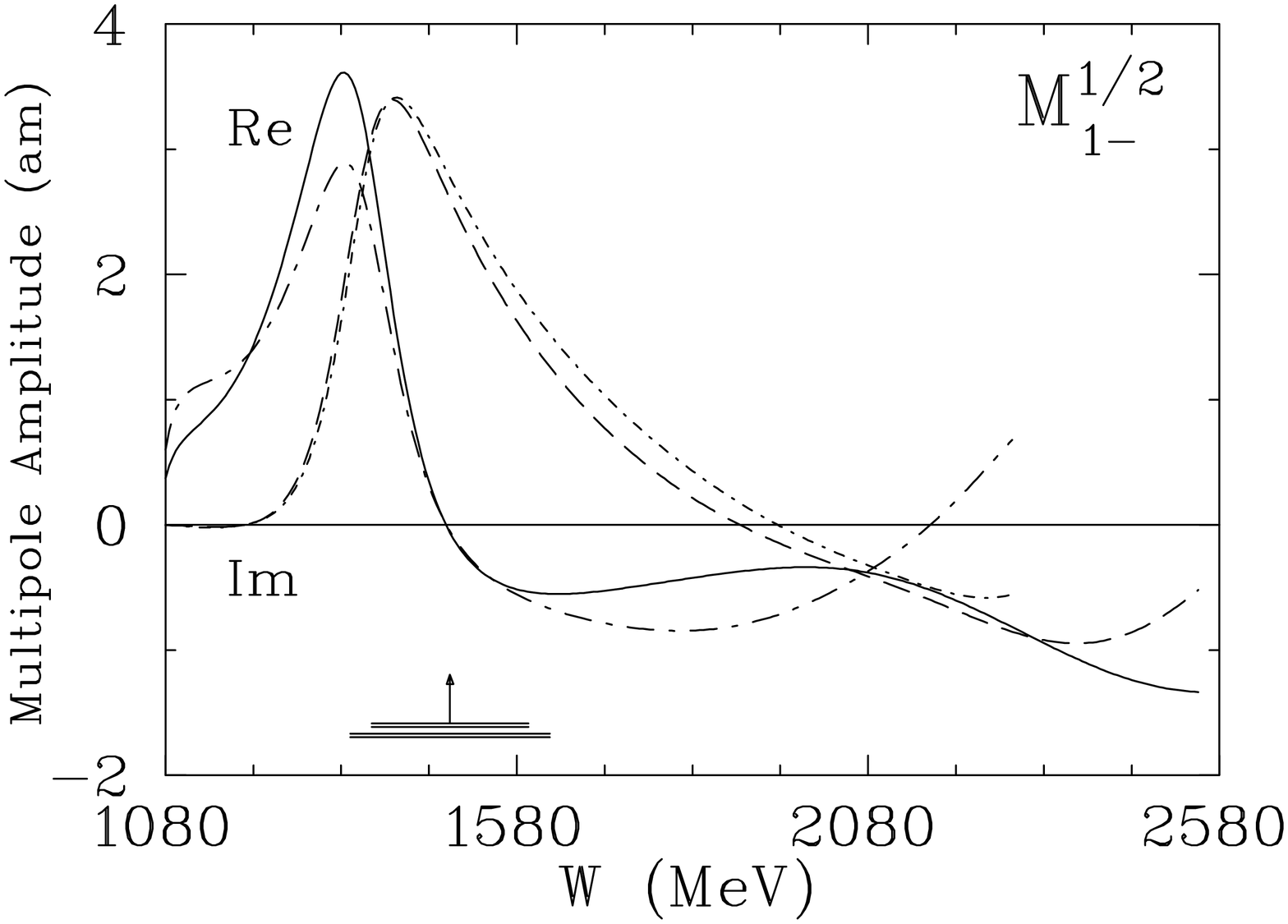}}
\centerline{
\includegraphics[height=0.45\textwidth, angle=90]{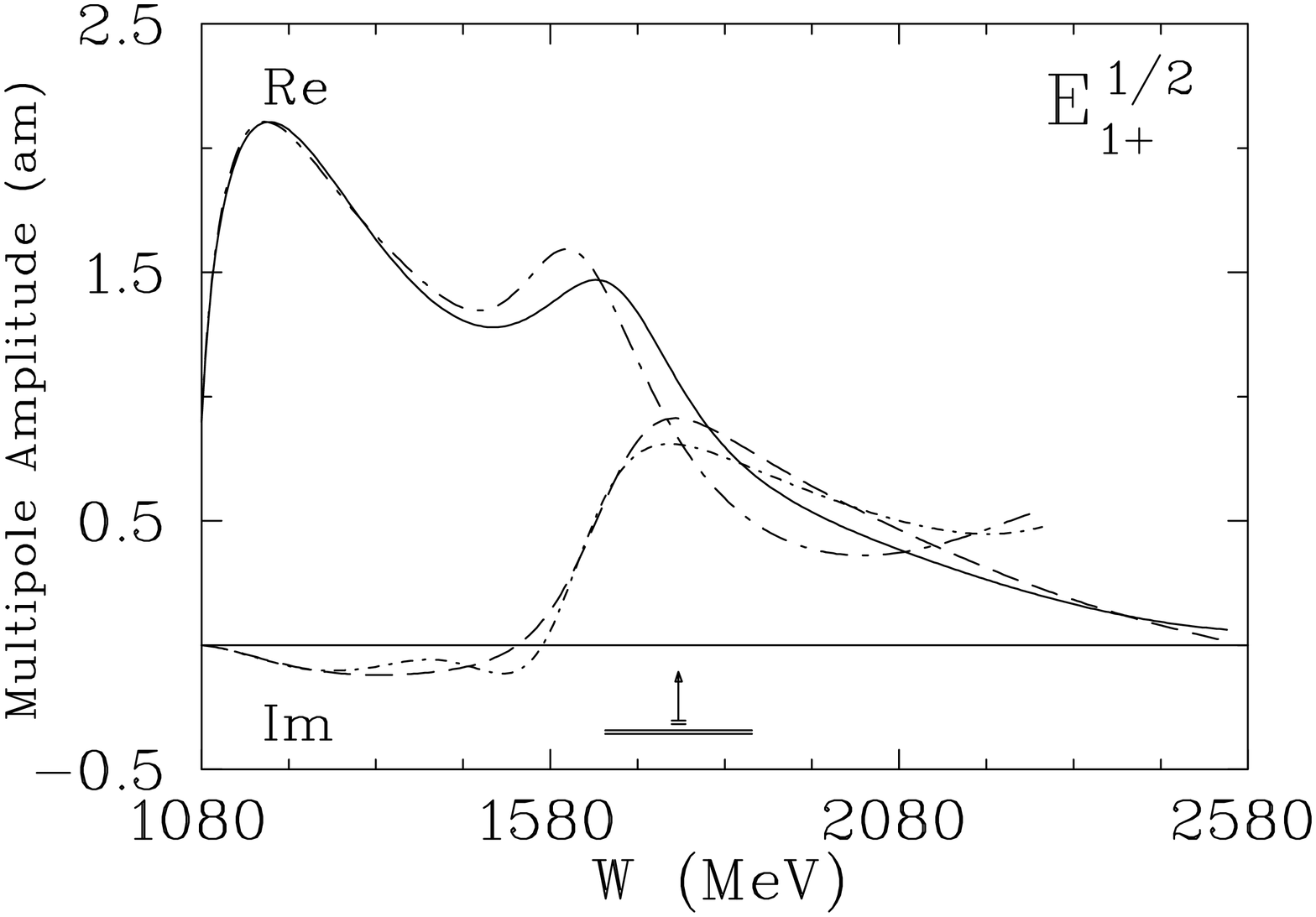}\hfill
\includegraphics[height=0.45\textwidth, angle=90]{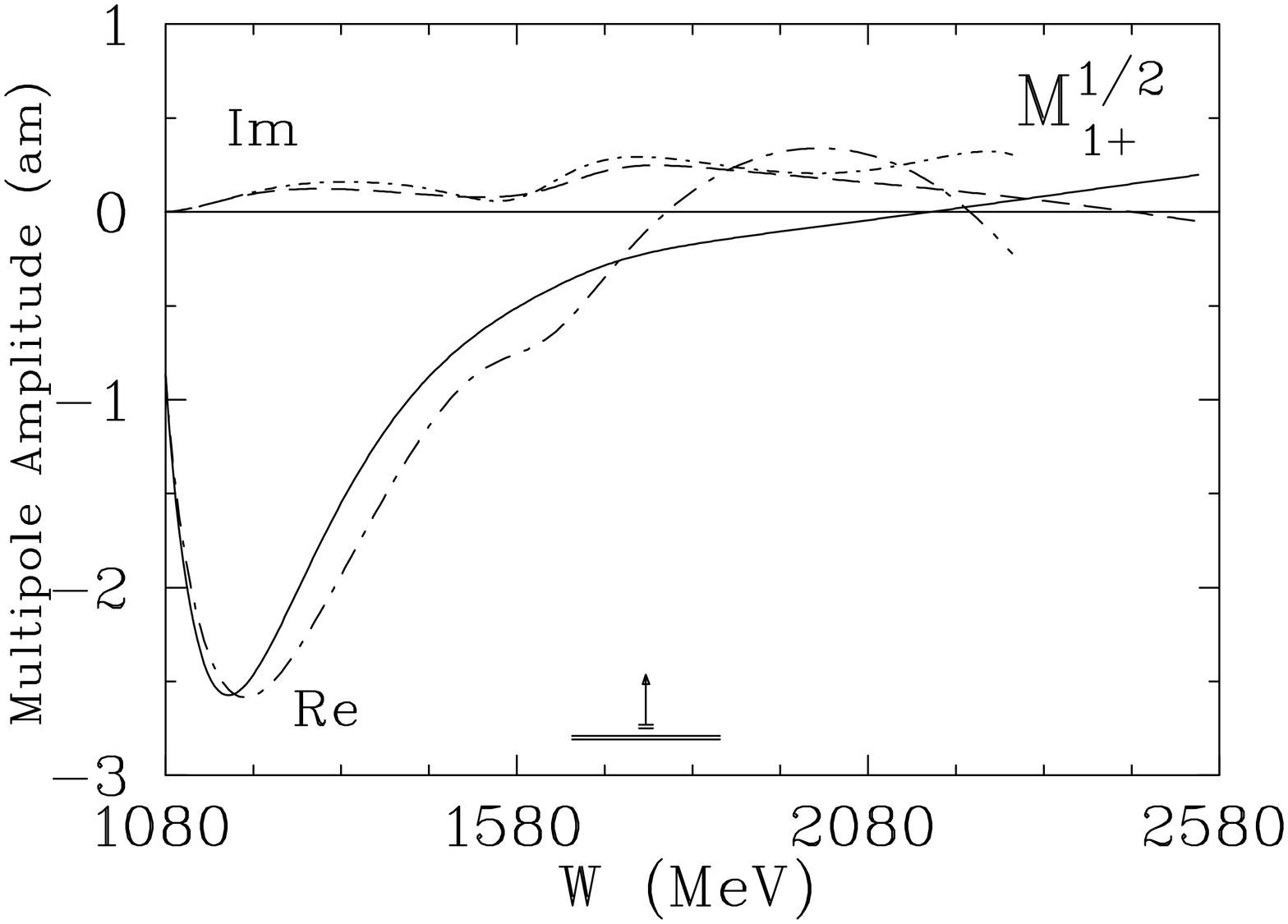}}
\centerline{
\includegraphics[height=0.45\textwidth, angle=90]{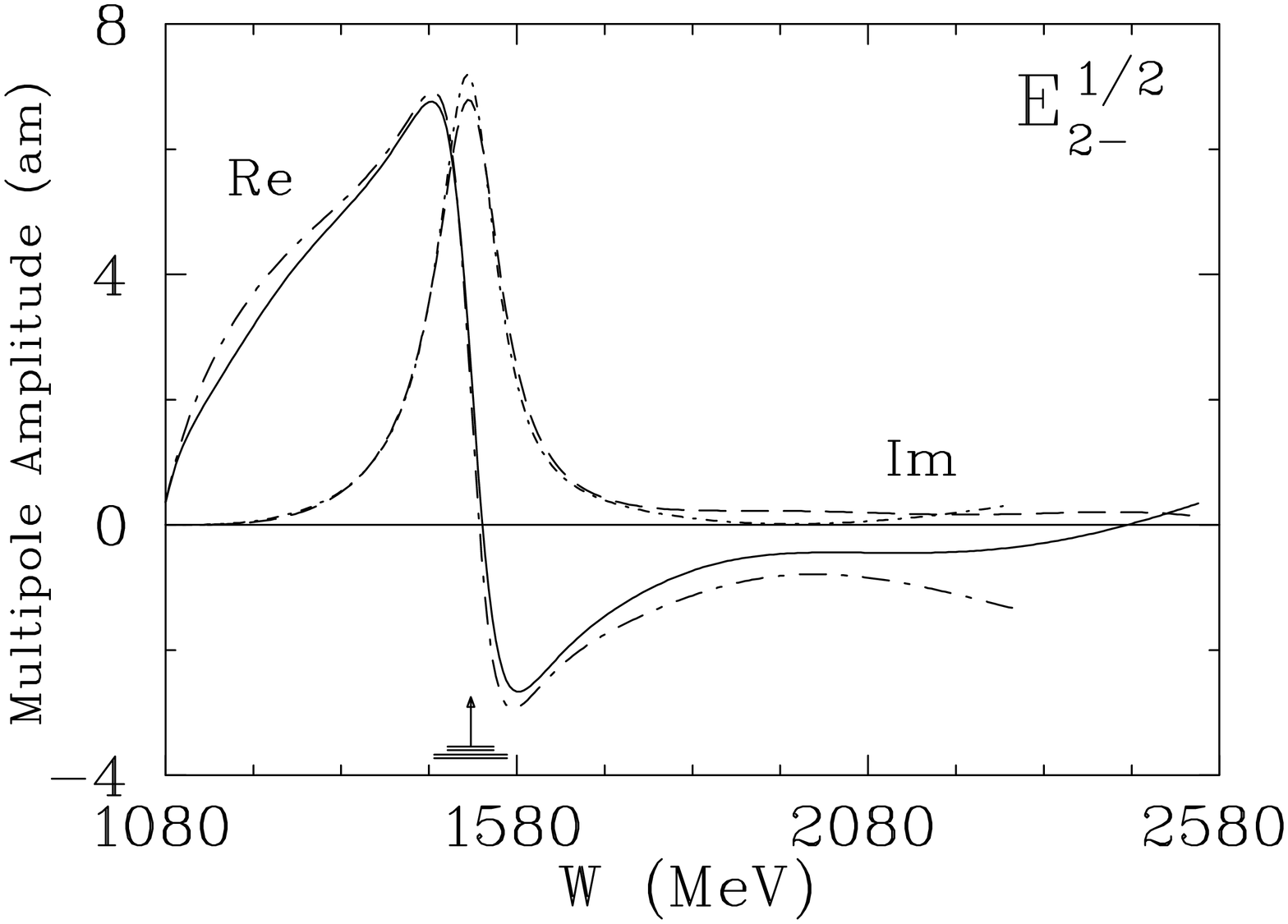}\hfill
\includegraphics[height=0.45\textwidth, angle=90]{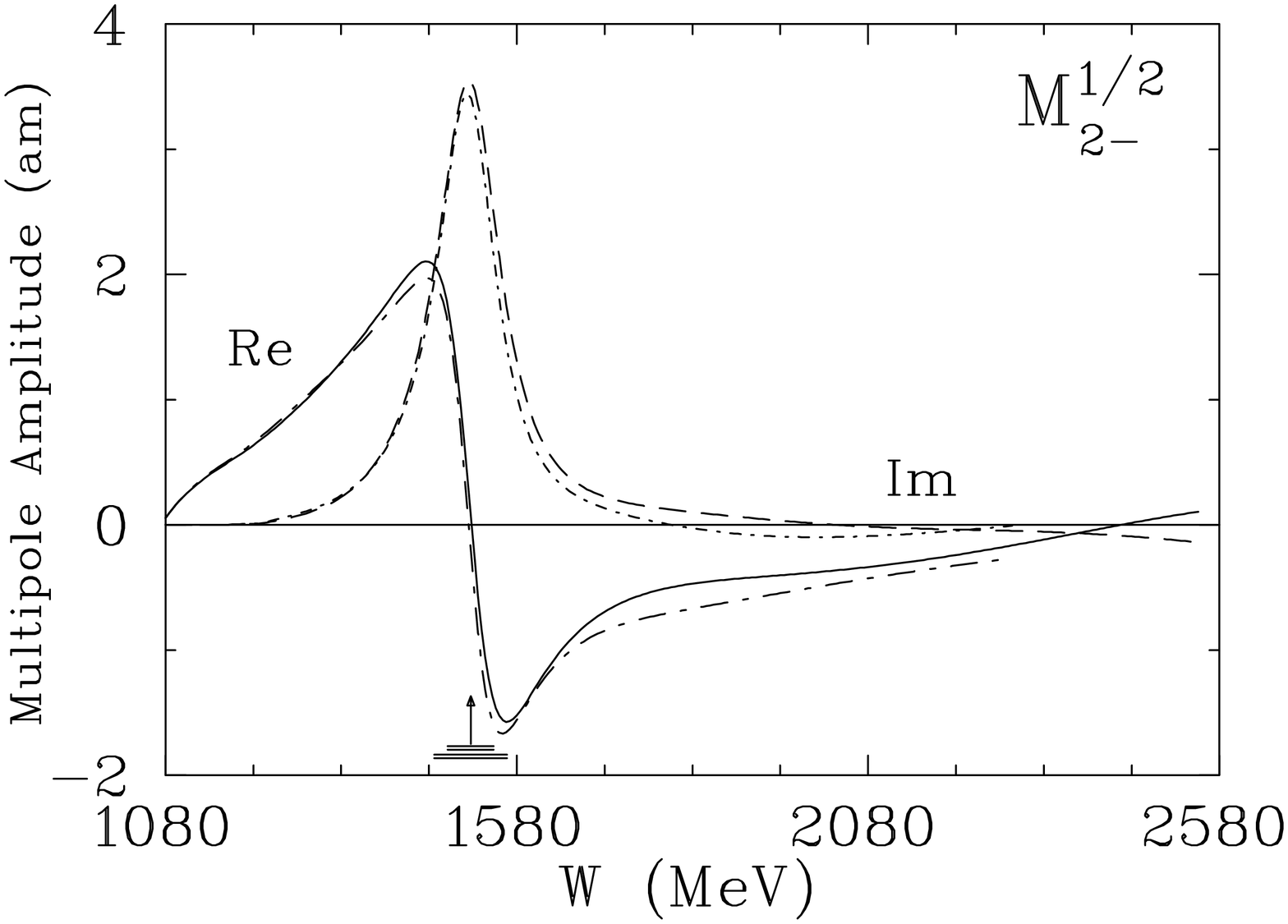}}
\centerline{
\includegraphics[height=0.45\textwidth, angle=90]{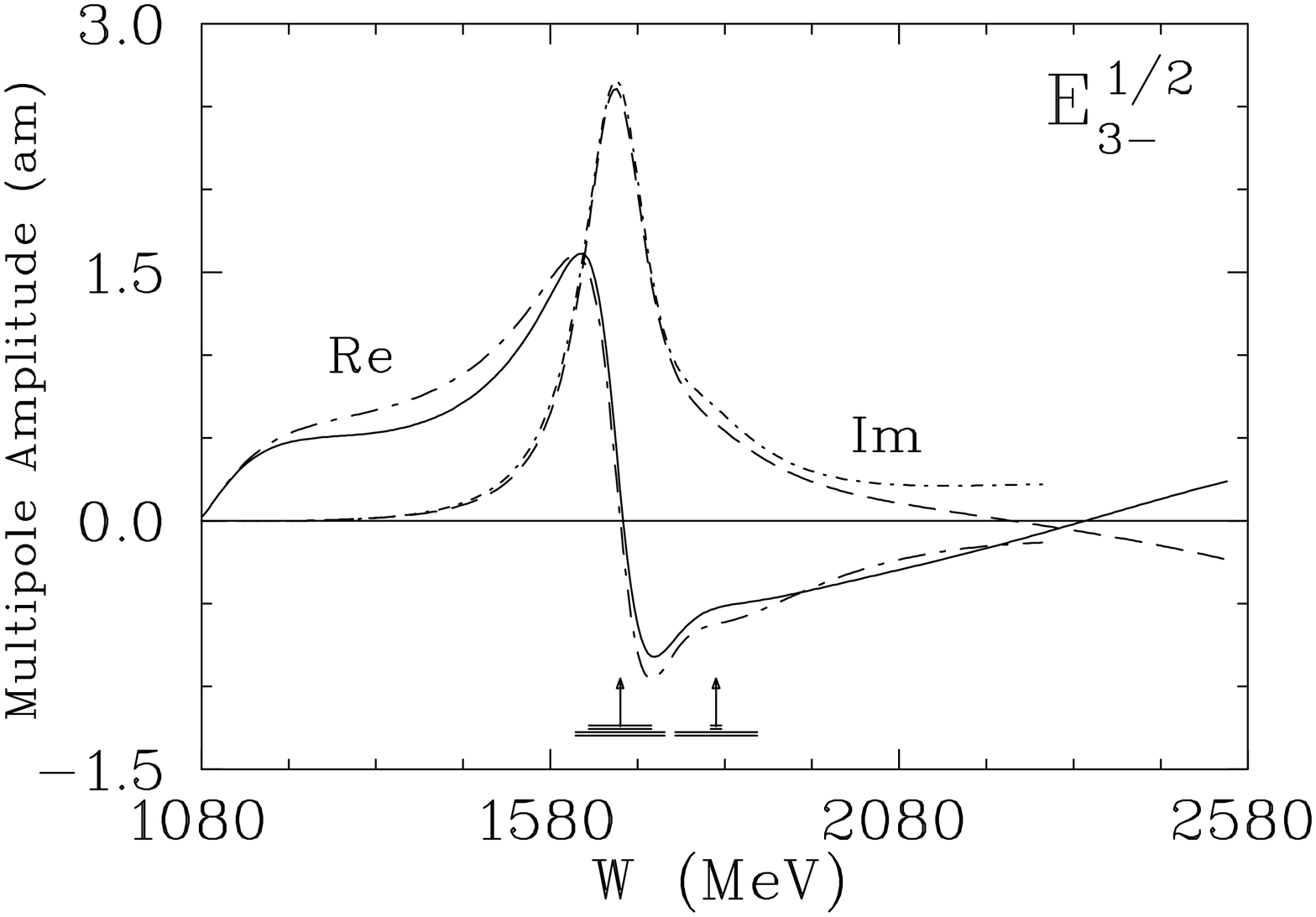}\hfill
\includegraphics[height=0.45\textwidth, angle=90]{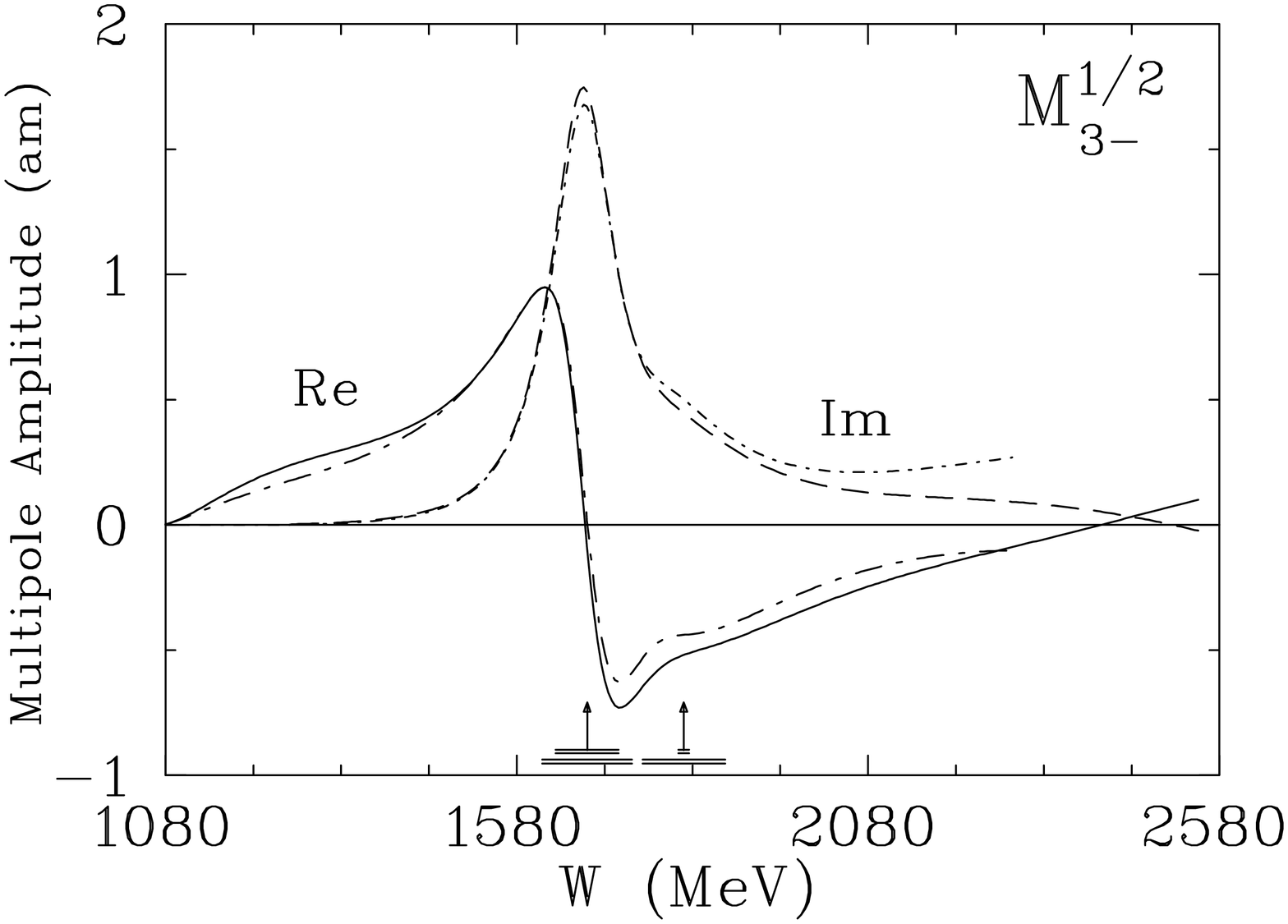}}
\caption{Multipole amplitudes from threshold to $E_{\gamma}$ 
         = 3~GeV for isospin $1/2$.  Solid (dashed) curves give the
         real (imaginary) parts of amplitudes corresponding to the
         SAID FA06 solution.  The previous SAID SM02
         solution~\protect\cite{sm02} is given by long dash-dotted
         (real part) and short dash-dotted (imaginary part)
         lines. Vertical arrows indicate $W_R$ and horizontal bars
         show full $\Gamma$ and partial widths for $\Gamma_{\pi N}$
         associated with the SAID $\pi N$ solution
         SP06~\protect\cite{sp06}.  \label{fig:g4}}
\end{figure*}
\begin{figure*}[th]
\centerline{
\includegraphics[height=0.45\textwidth, angle=90]{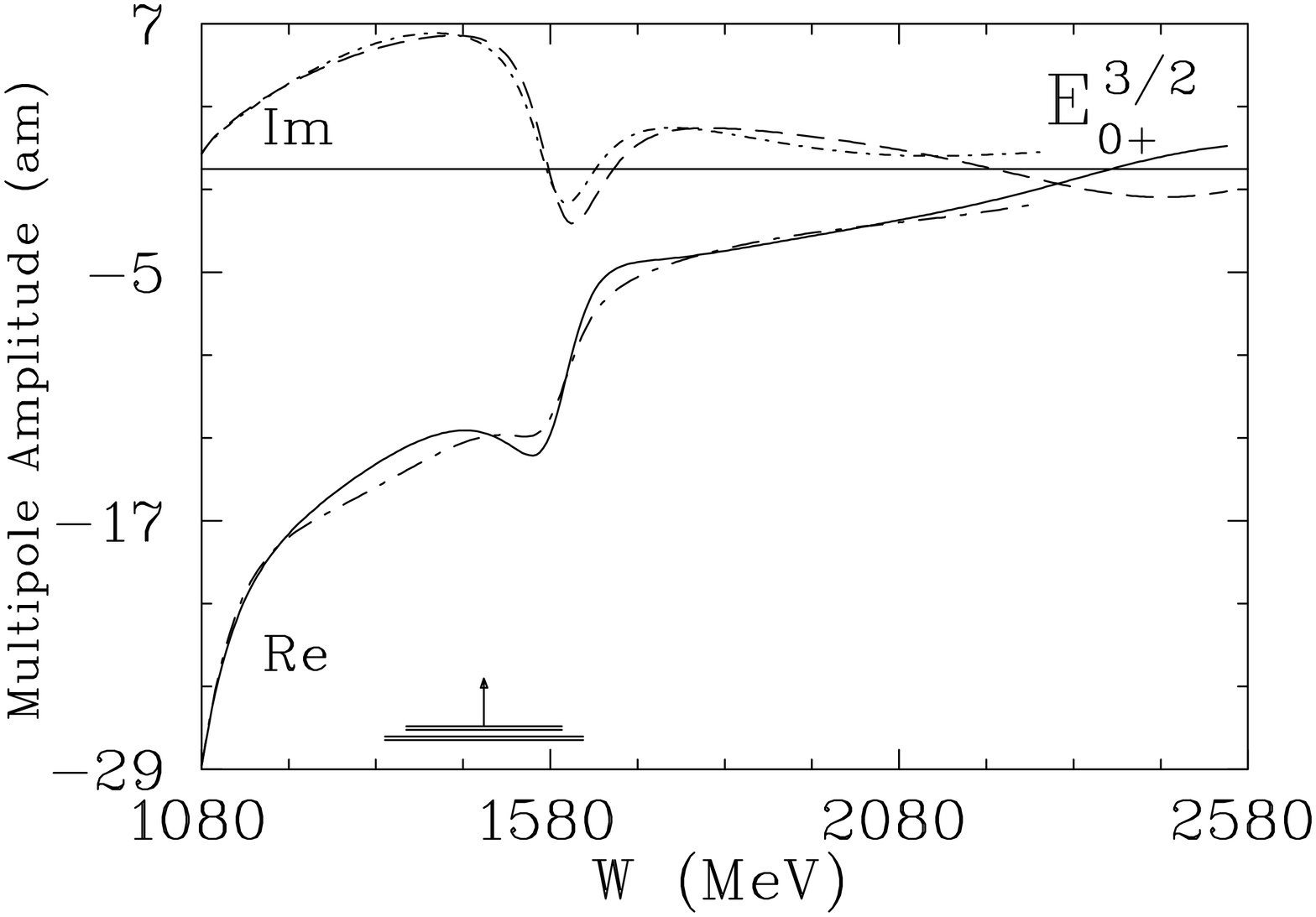}\hfill
\includegraphics[height=0.45\textwidth, angle=90]{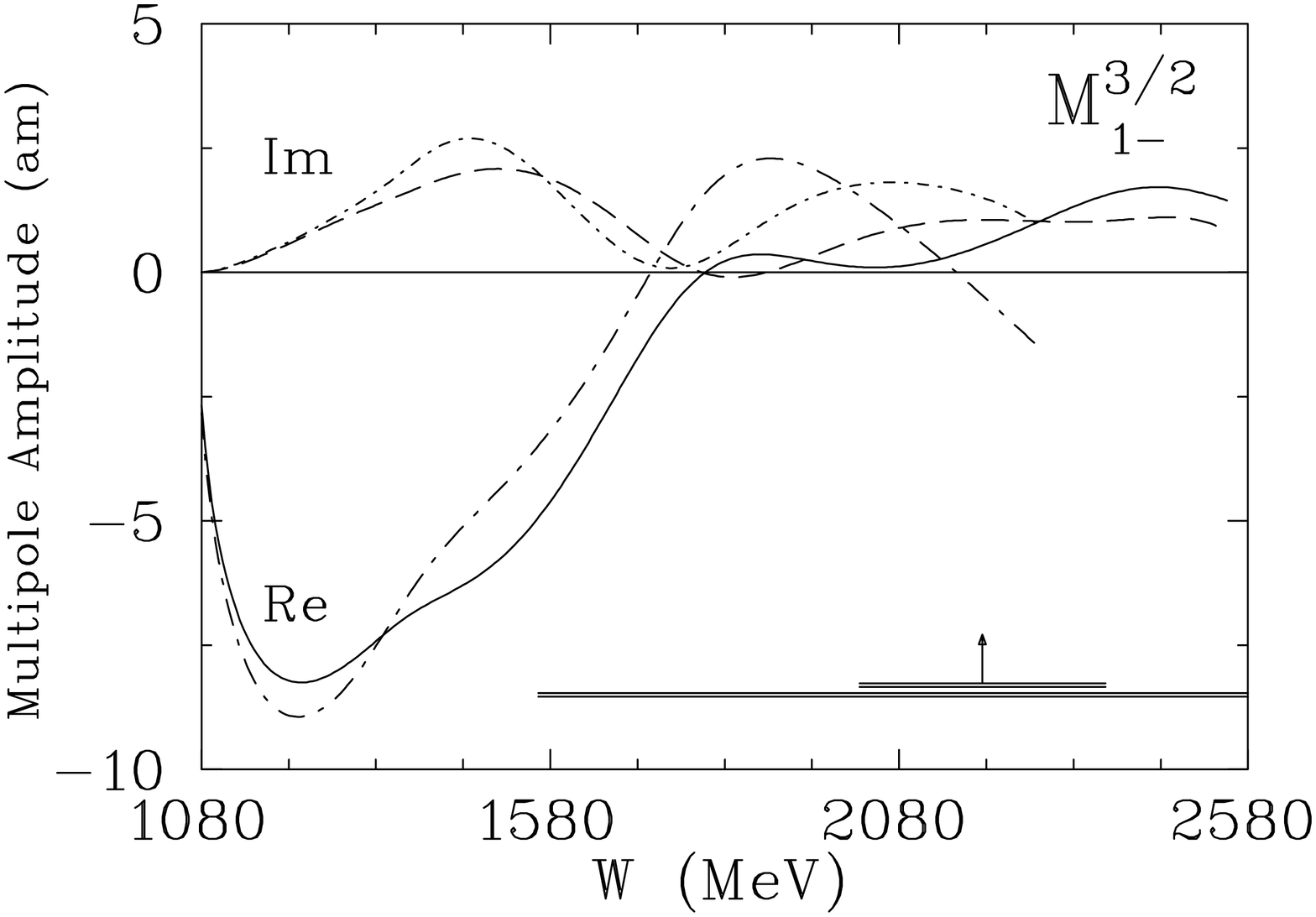}}
\centerline{
\includegraphics[height=0.45\textwidth, angle=90]{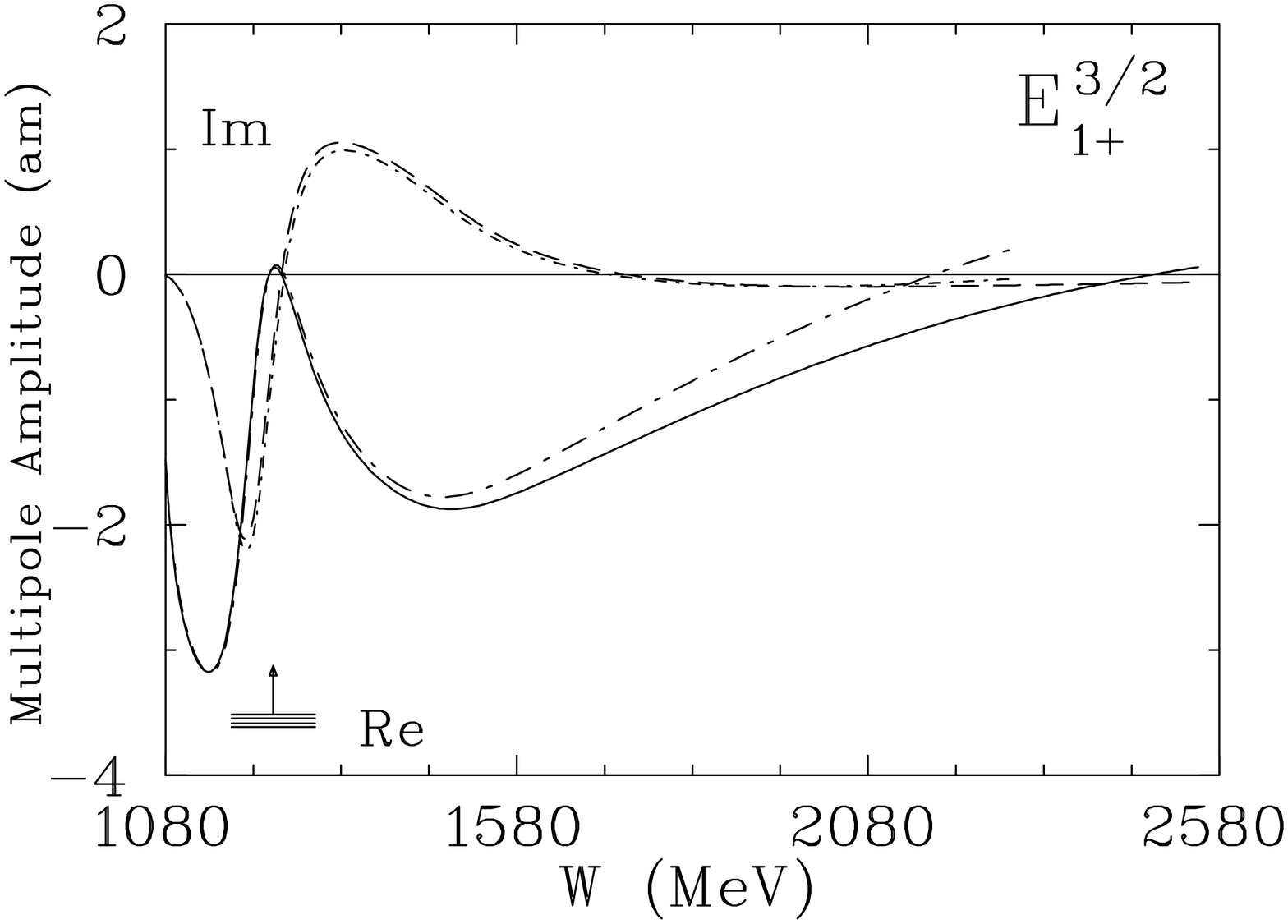}\hfill
\includegraphics[height=0.45\textwidth, angle=90]{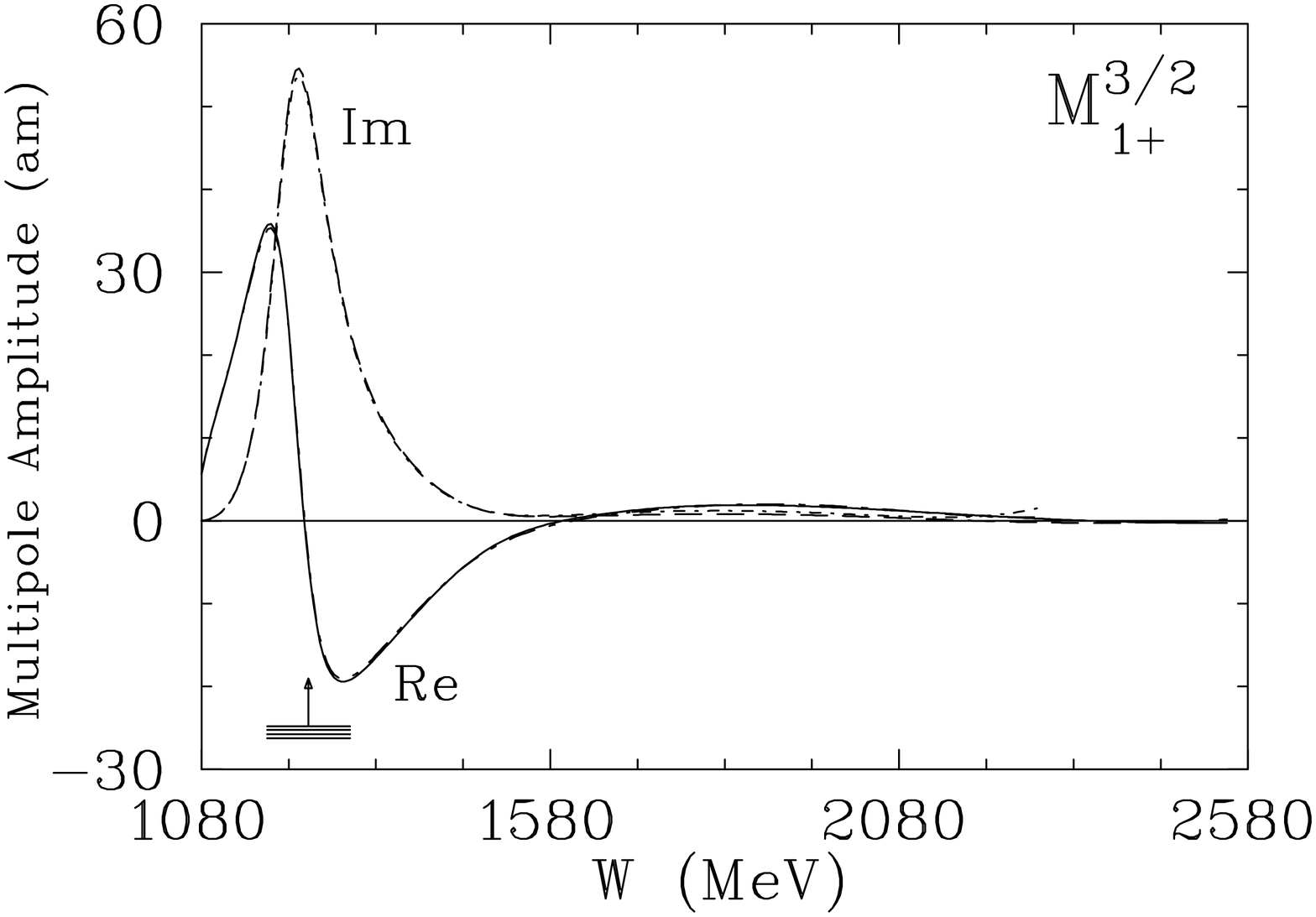}}
\centerline{
\includegraphics[height=0.45\textwidth, angle=90]{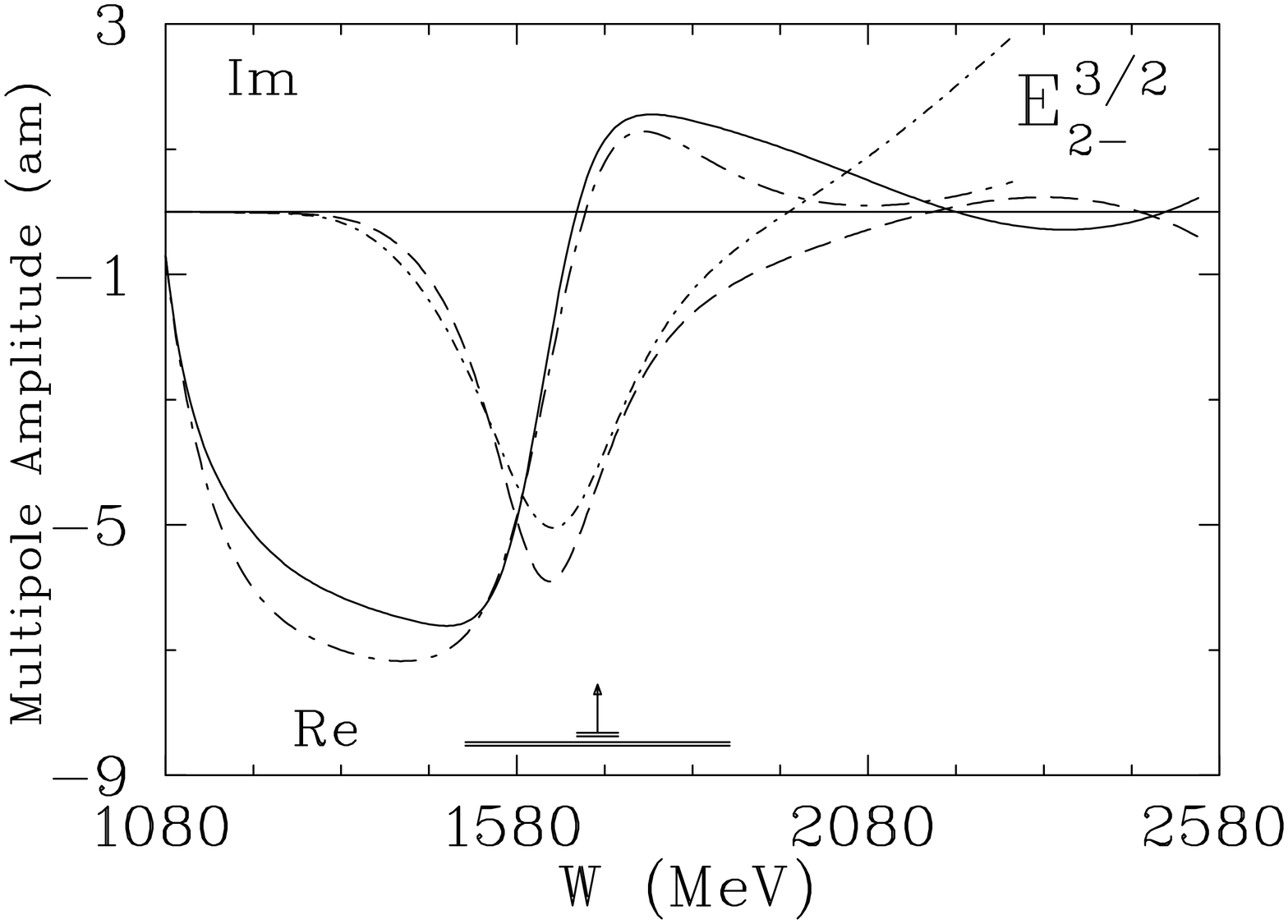}\hfill
\includegraphics[height=0.45\textwidth, angle=90]{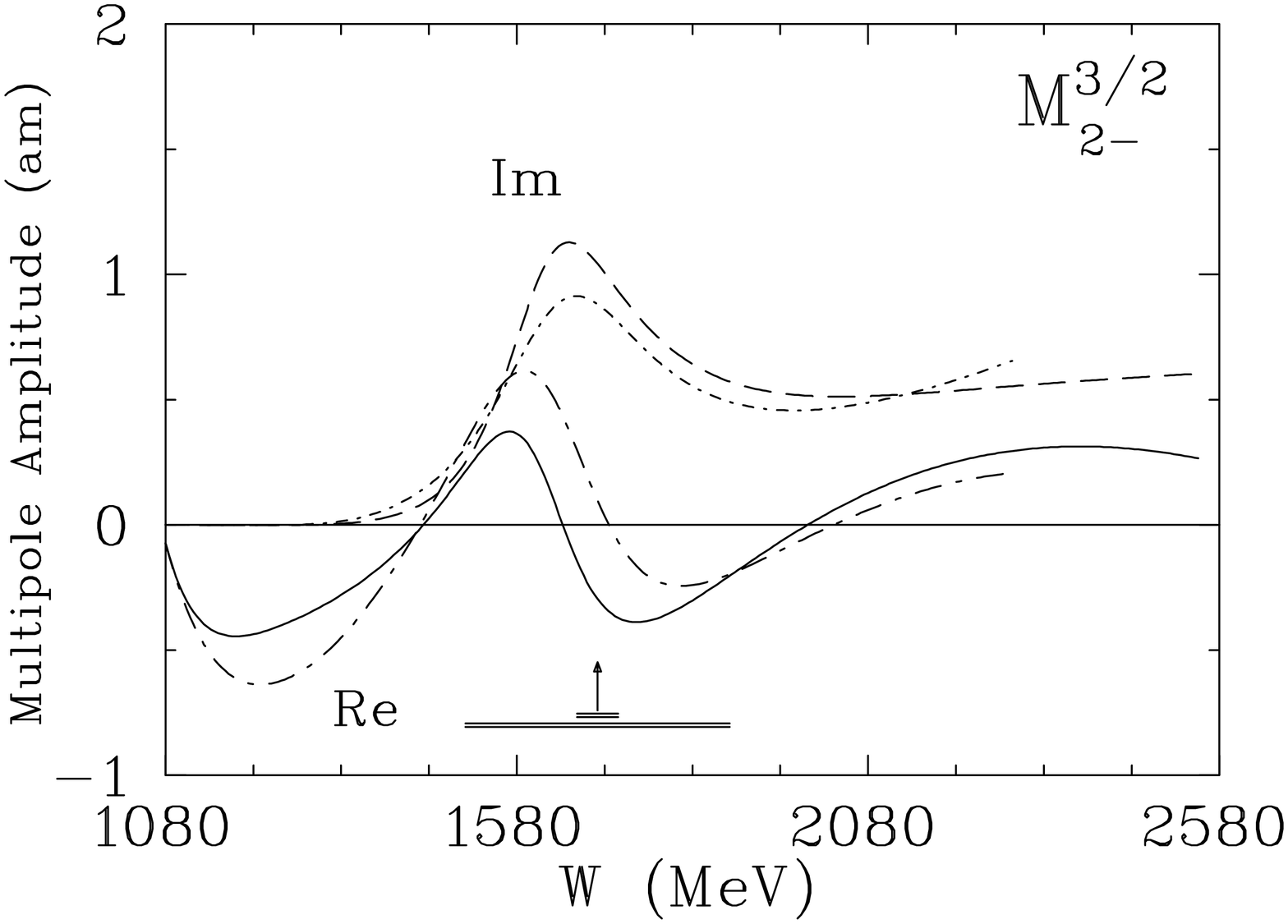}}
\centerline{
\includegraphics[height=0.45\textwidth, angle=90]{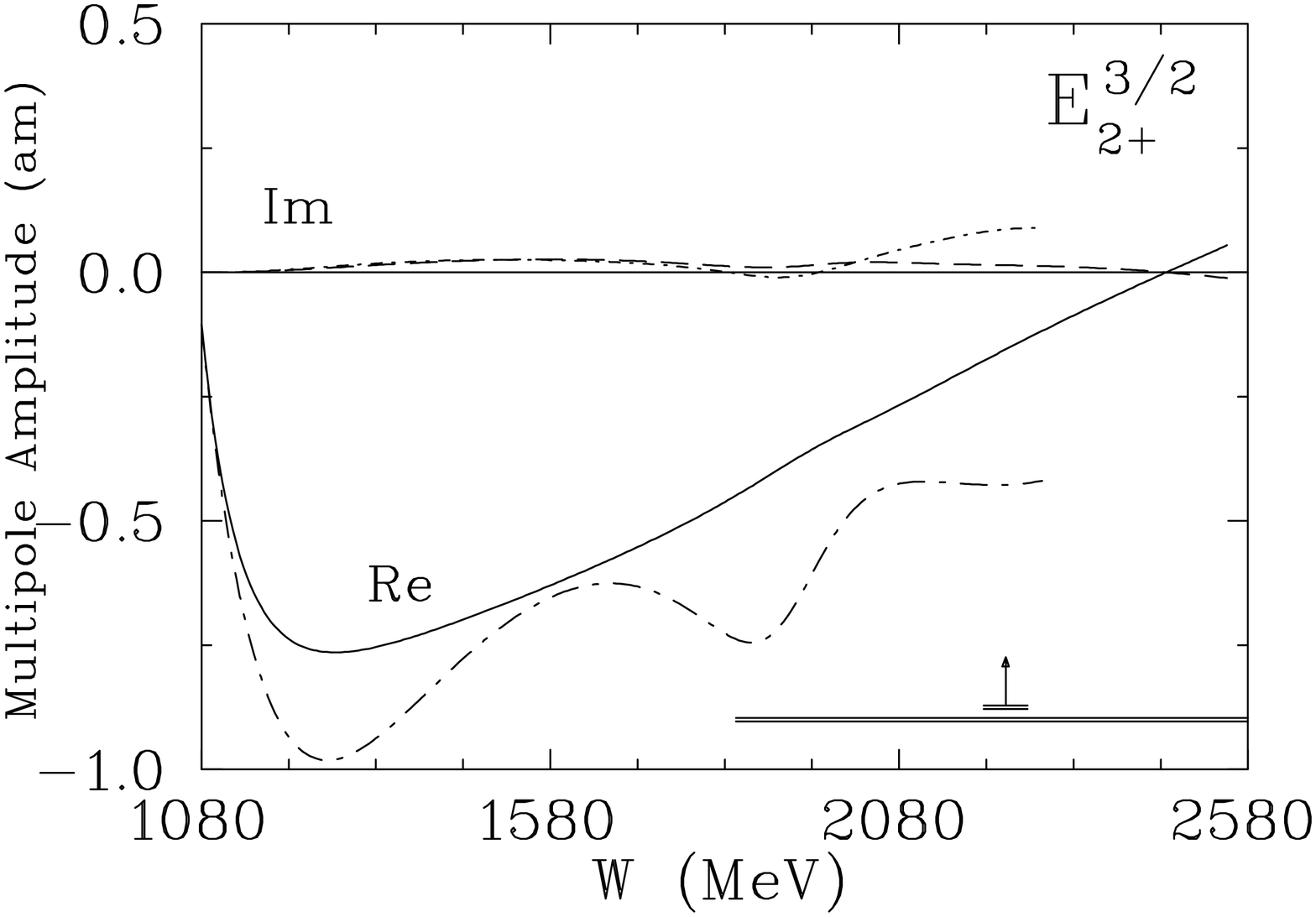}\hfill
\includegraphics[height=0.45\textwidth, angle=90]{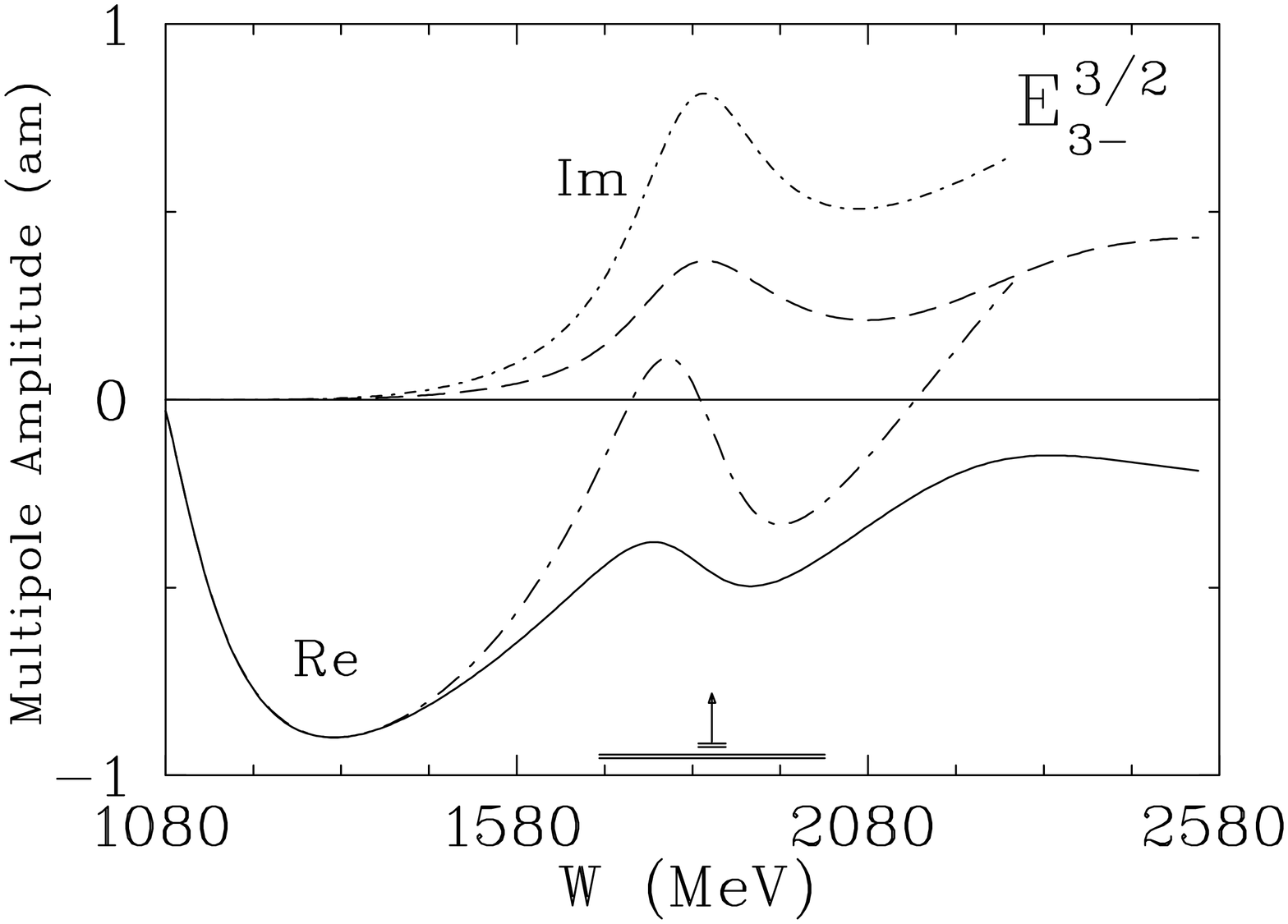}}
\caption{Multipole amplitudes from threshold to $E_{\gamma}$ 
         = 3~GeV for isospin $3/2$.  Notation as in
         Fig.~\protect\ref{fig:g4}. \label{fig:g5}}
\end{figure*}
\begin{figure*}[th]
\centerline{
\includegraphics[height=0.45\textwidth, angle=90]{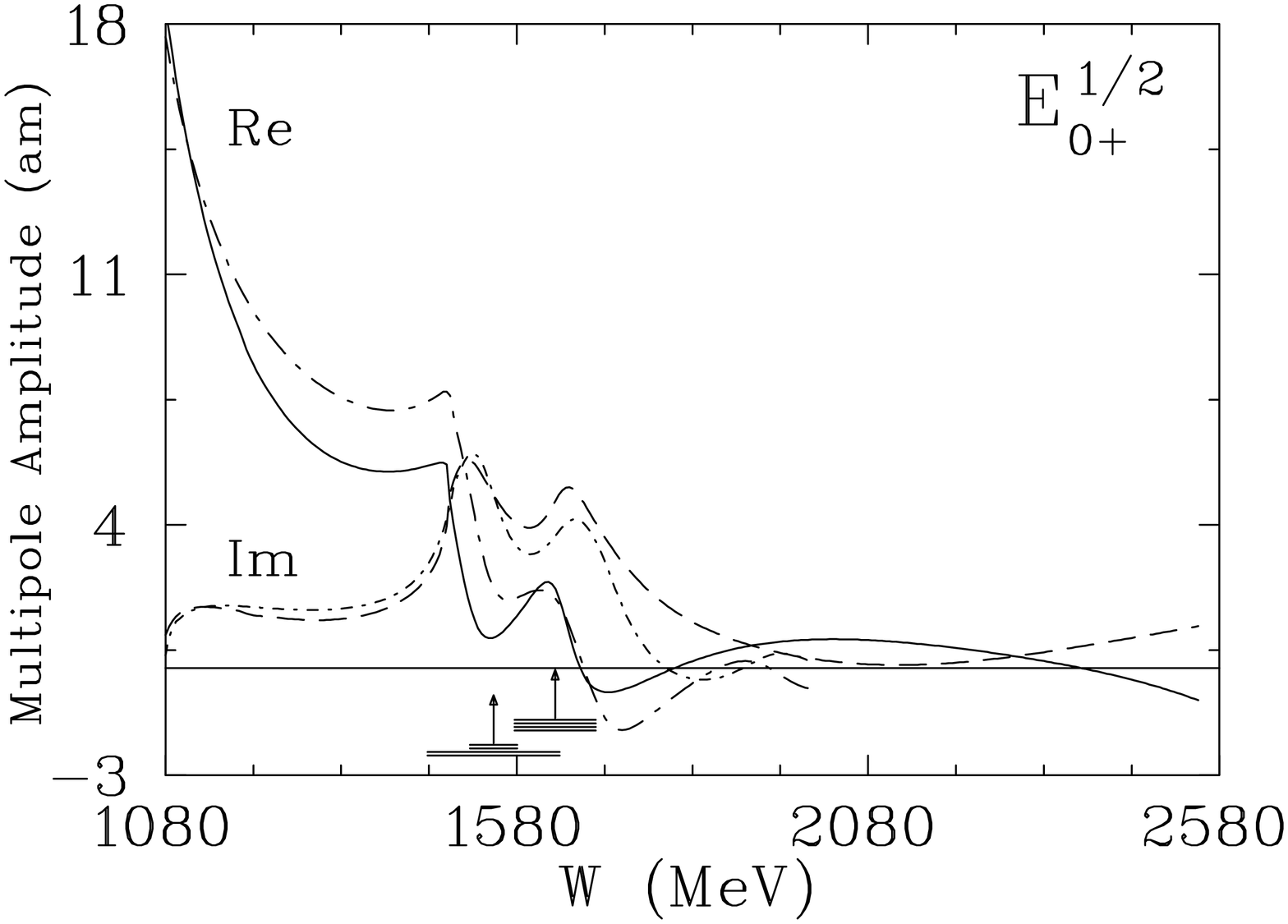}\hfill
\includegraphics[height=0.45\textwidth, angle=90]{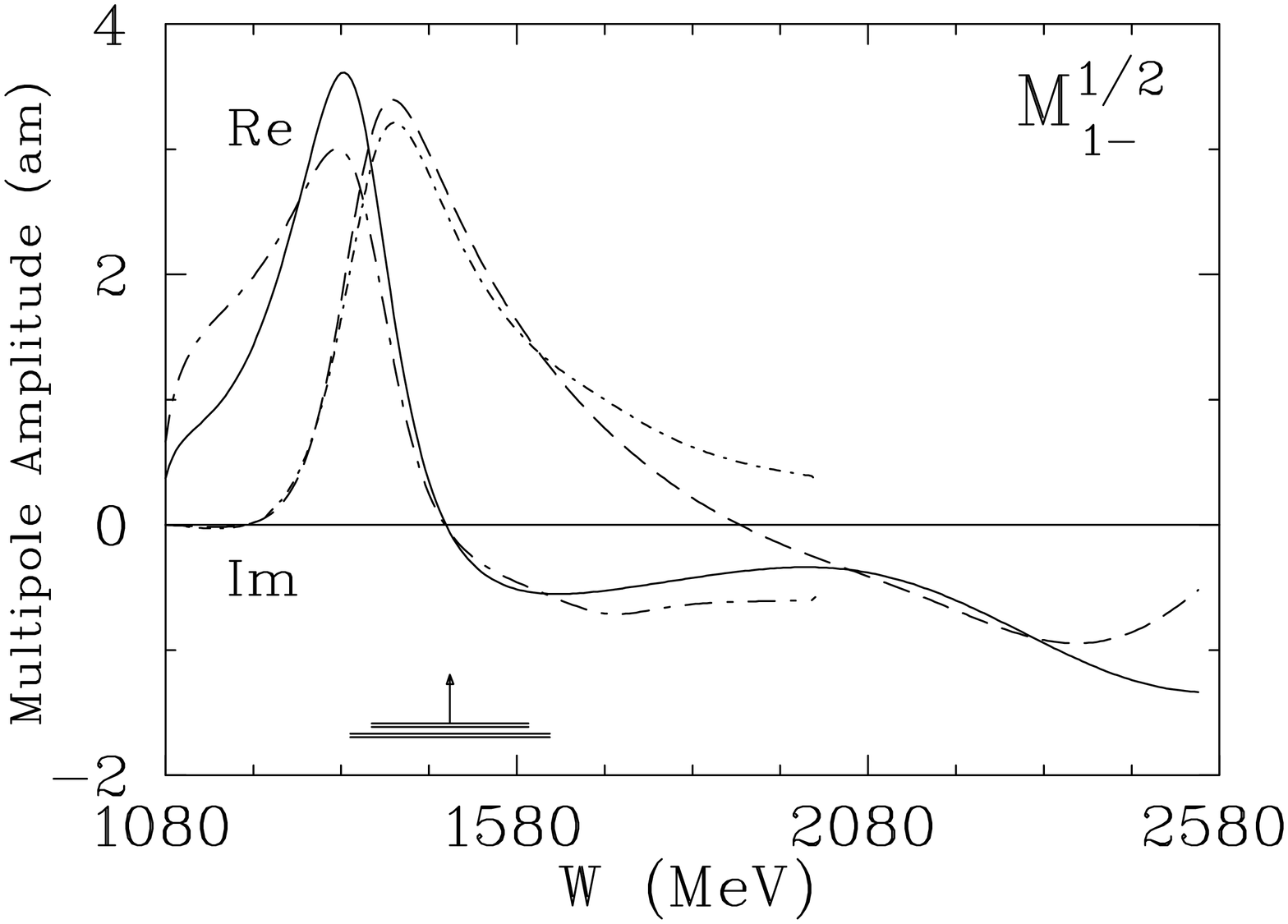}}
\centerline{
\includegraphics[height=0.45\textwidth, angle=90]{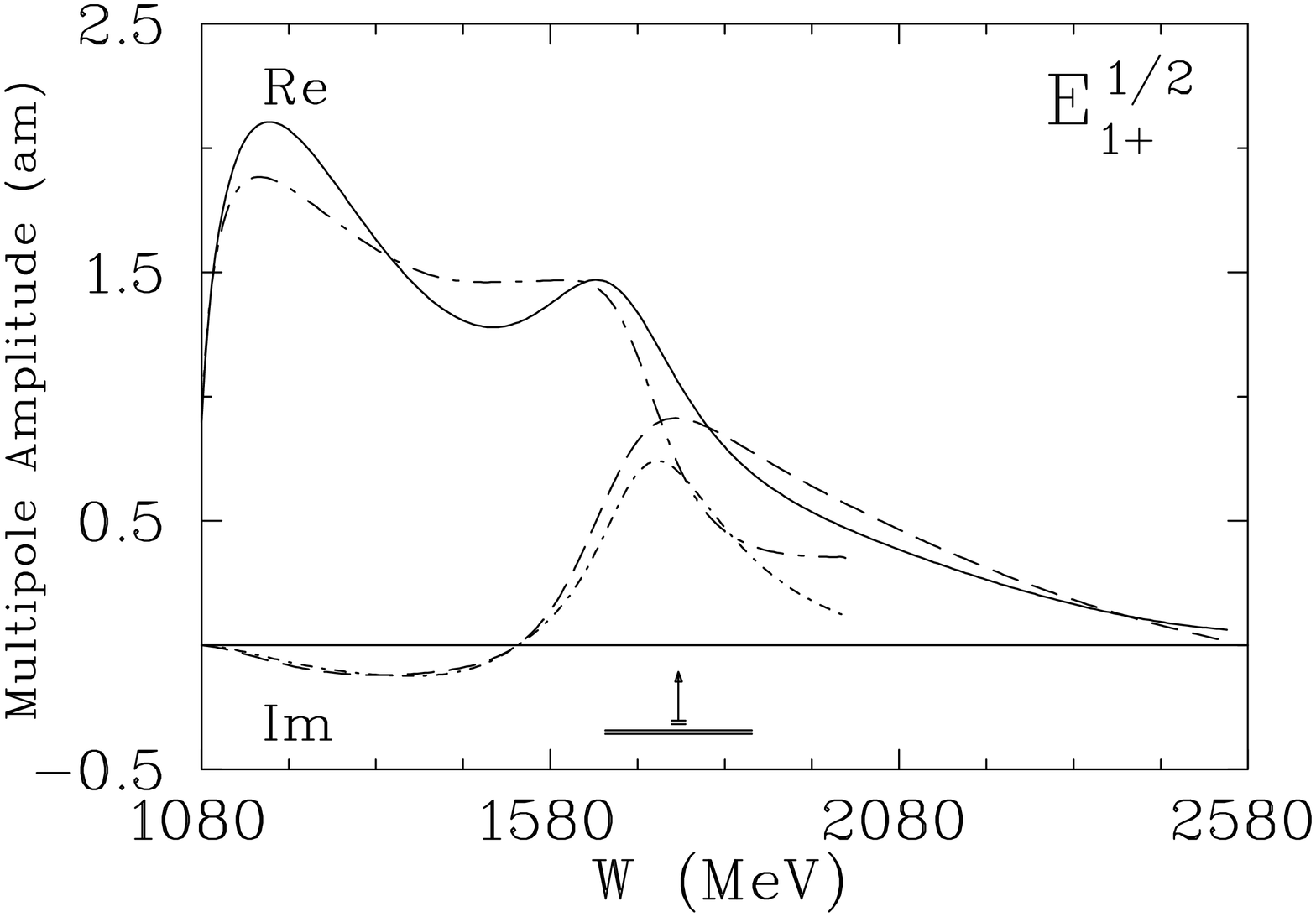}\hfill
\includegraphics[height=0.45\textwidth, angle=90]{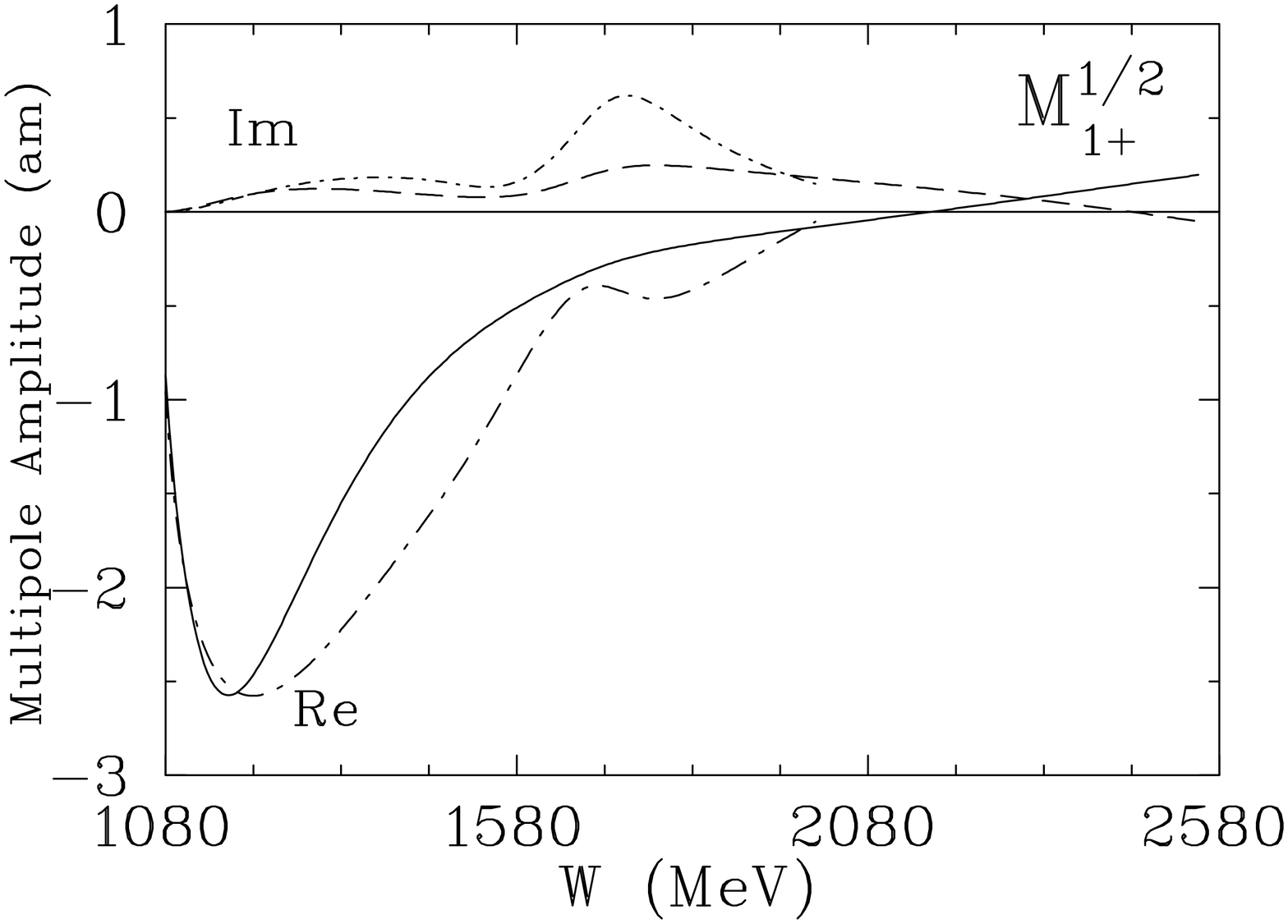}}
\centerline{
\includegraphics[height=0.45\textwidth, angle=90]{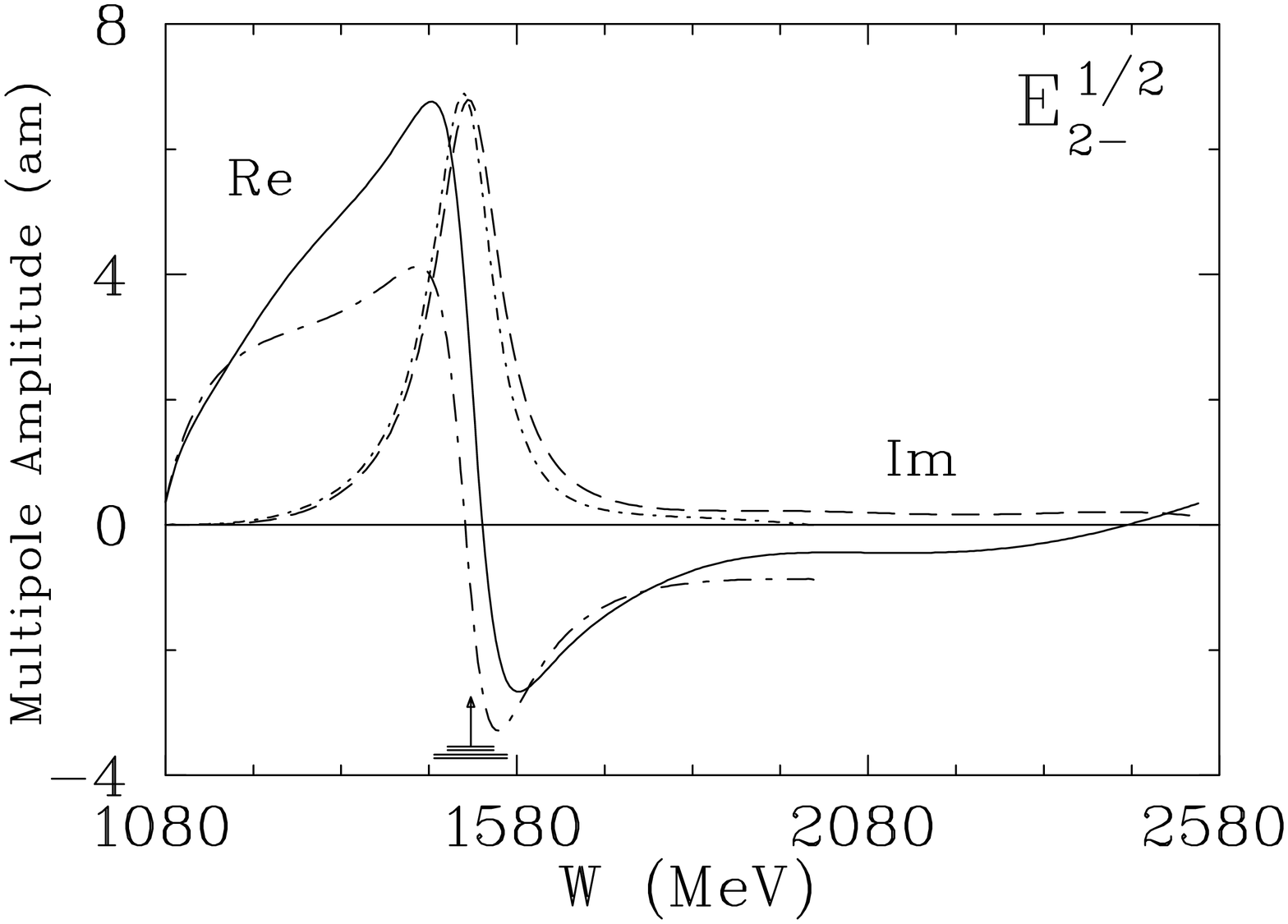}\hfill
\includegraphics[height=0.45\textwidth, angle=90]{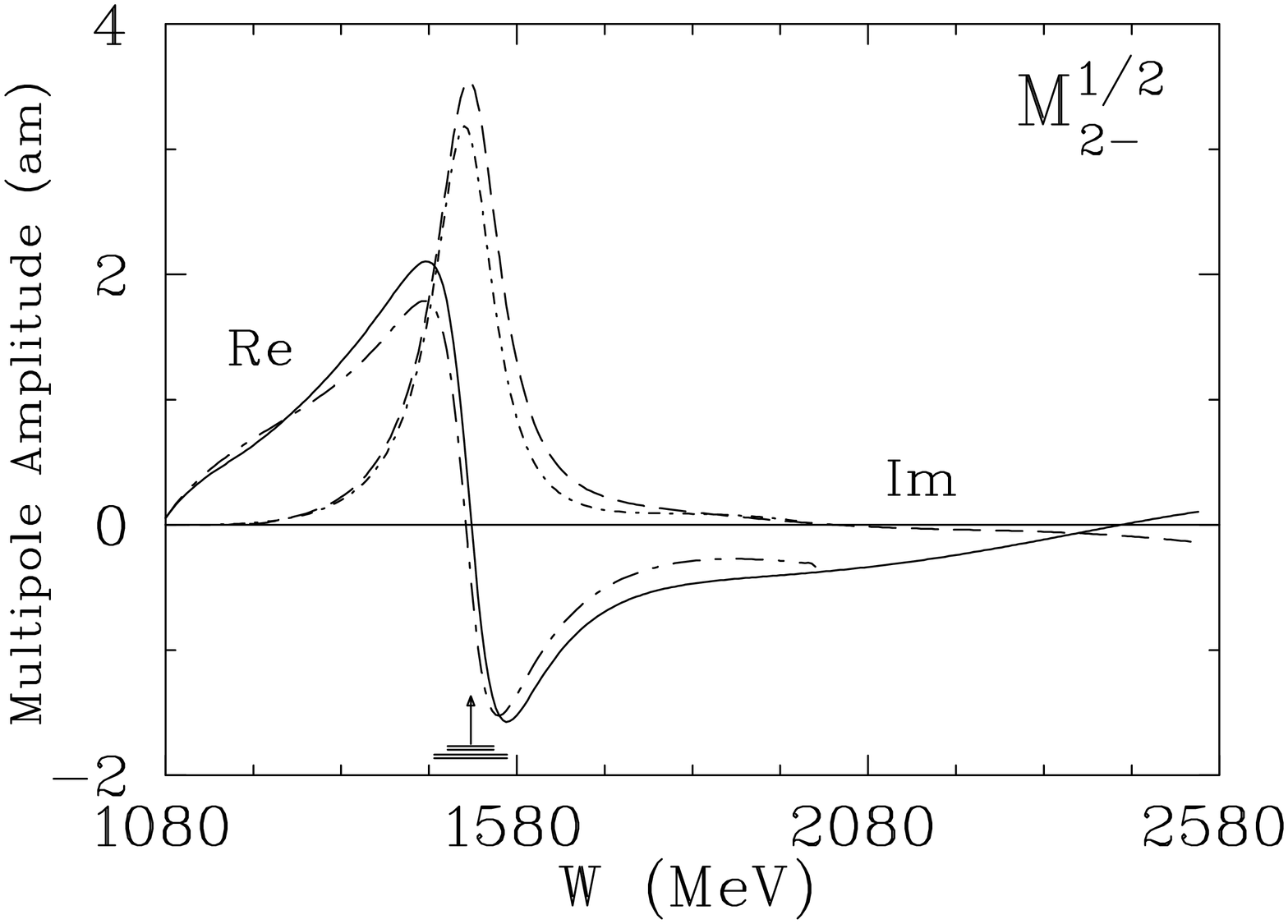}}
\centerline{
\includegraphics[height=0.45\textwidth, angle=90]{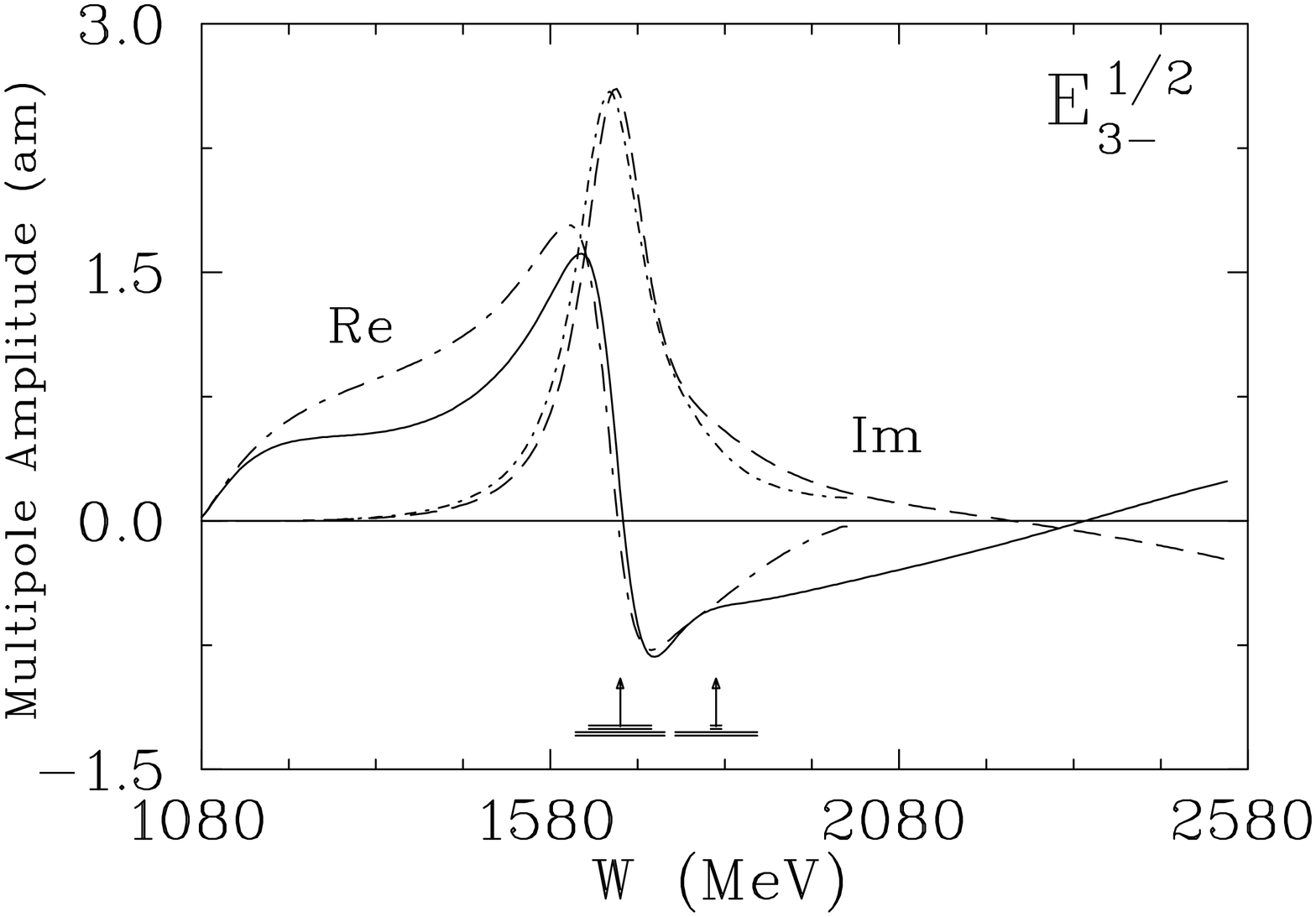}\hfill
\includegraphics[height=0.45\textwidth, angle=90]{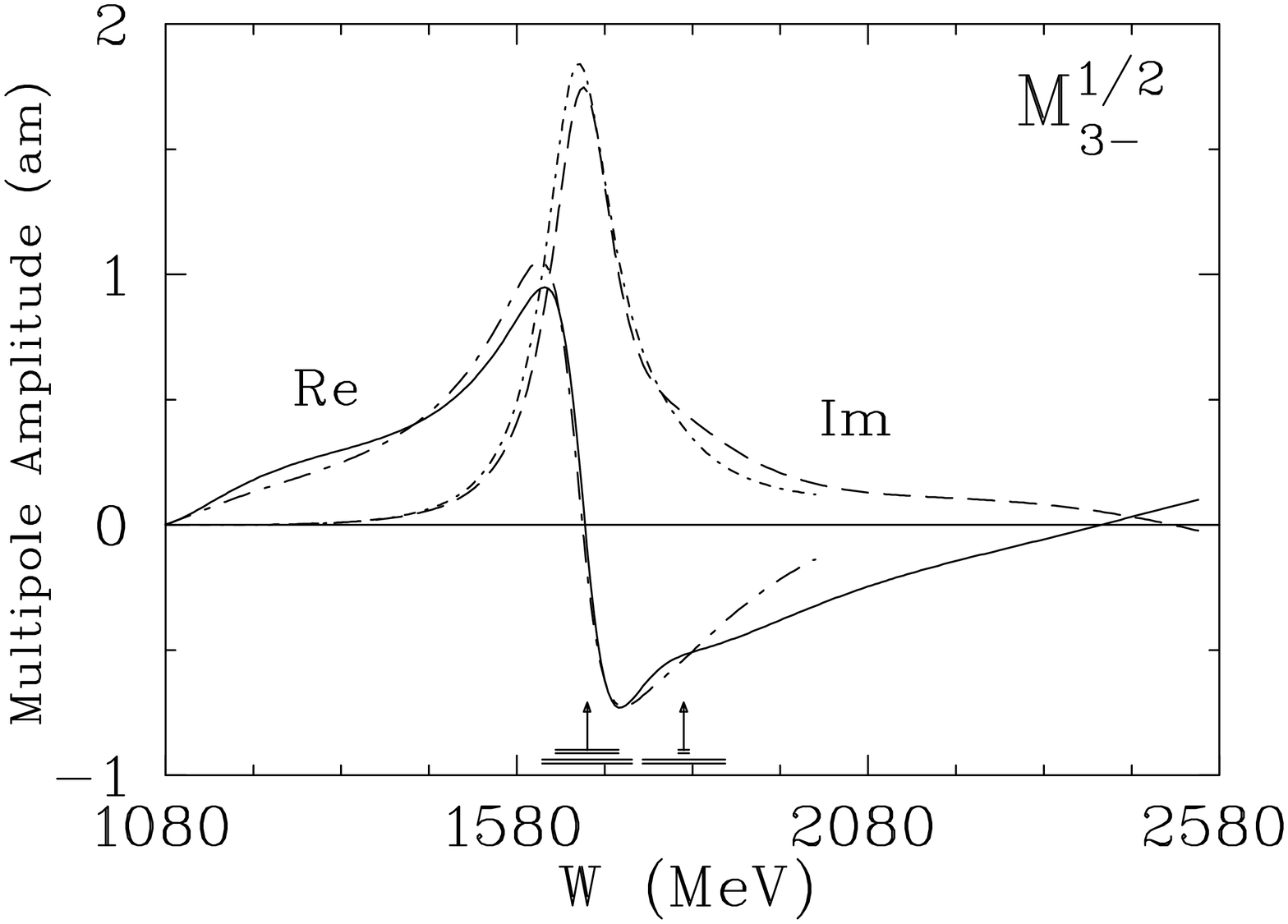}}
\caption{Multipole amplitudes from threshold to $E_{\gamma}$ 
         = 3~GeV for isospin $1/2$.  Solid (dashed) curves give the
         real (imaginary) parts of amplitudes corresponding to the
         SAID FA06 solution.  The MAID05 solution~\protect\cite{maid}
         is given by long dash-dotted (real part) and short
         dash-dotted (imaginary part) lines. \label{fig:g6}}
\end{figure*}
\begin{figure*}[th]
\centerline{
\includegraphics[height=0.45\textwidth, angle=90]{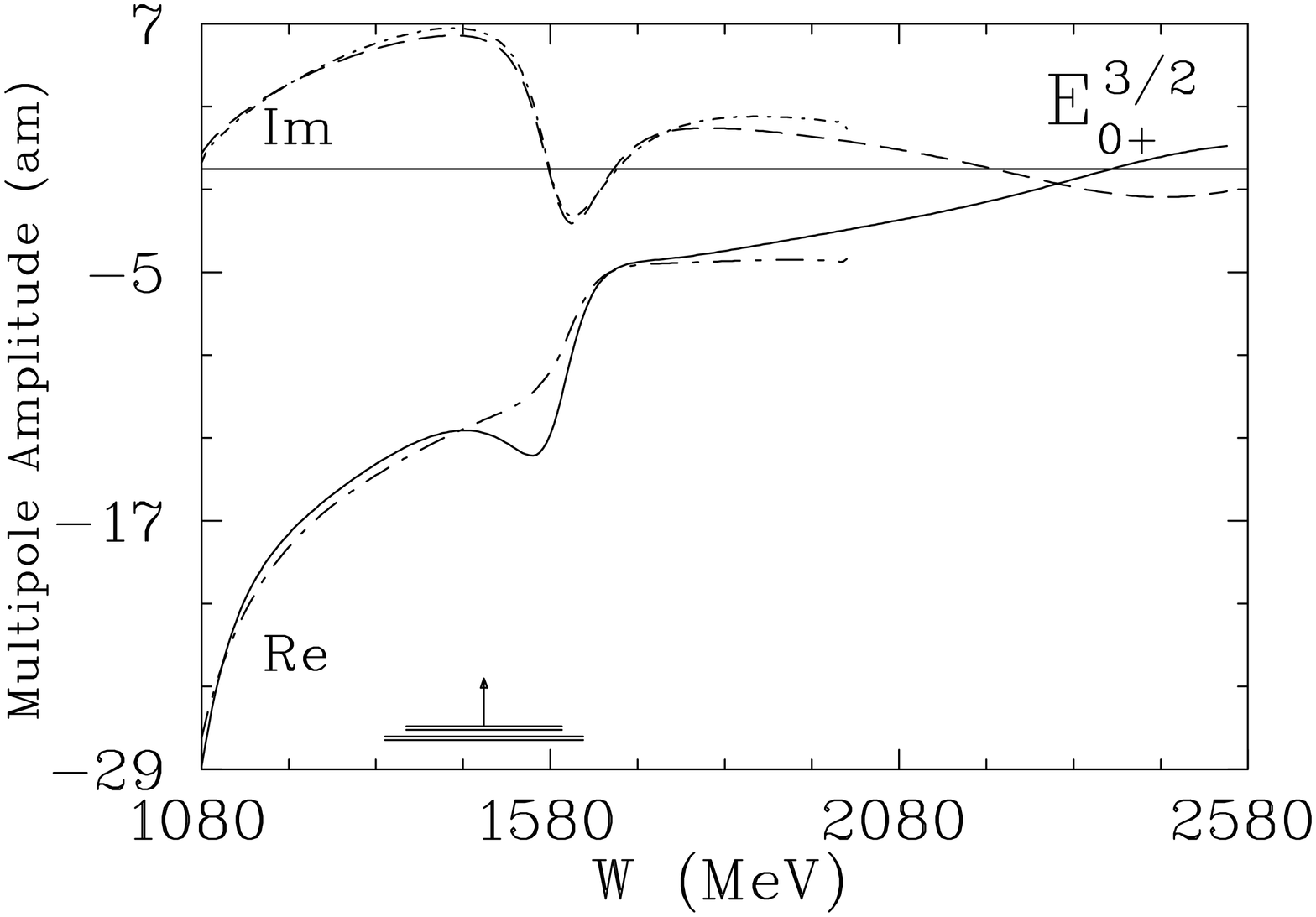}\hfill
\includegraphics[height=0.45\textwidth, angle=90]{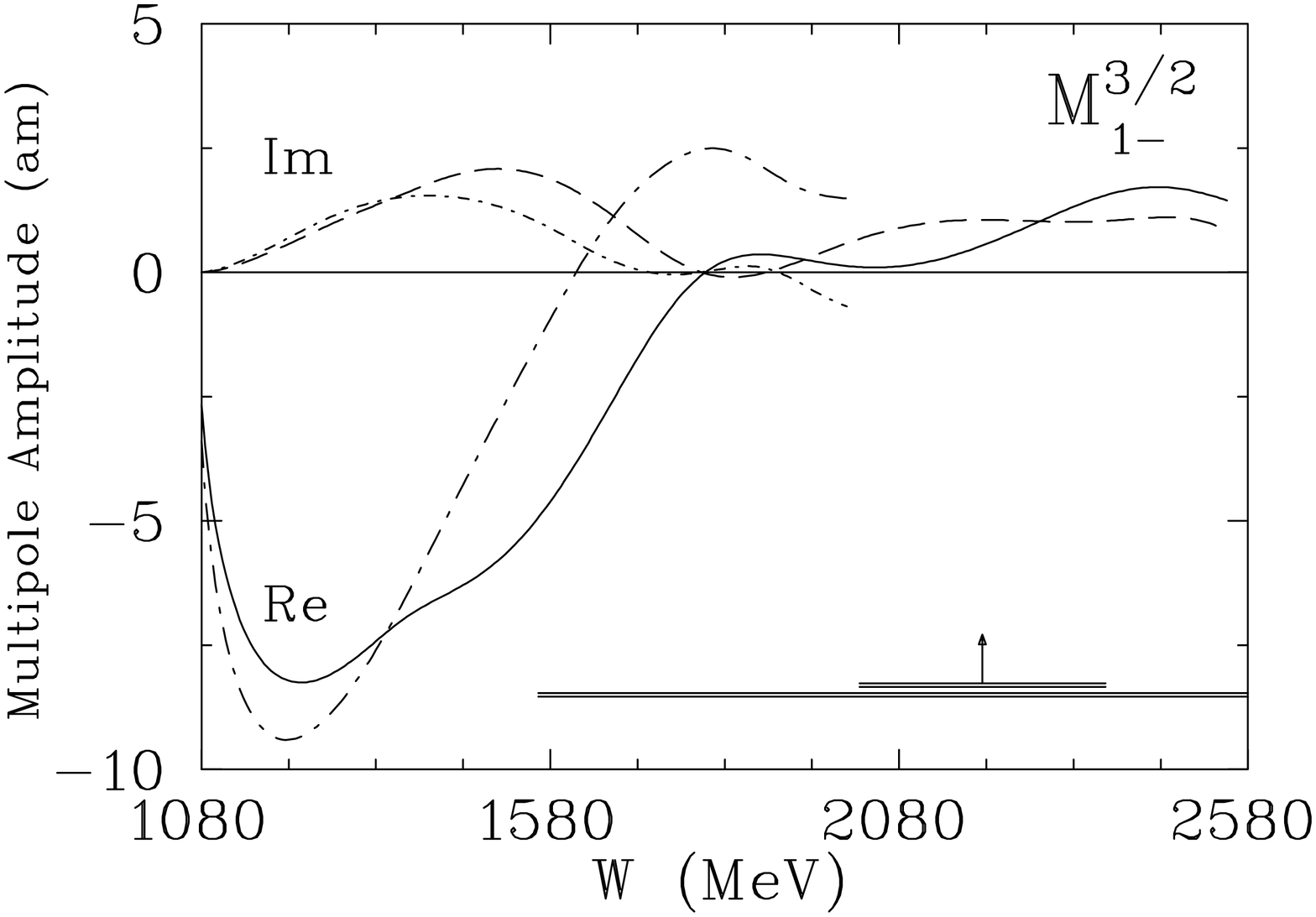}}
\centerline{
\includegraphics[height=0.45\textwidth, angle=90]{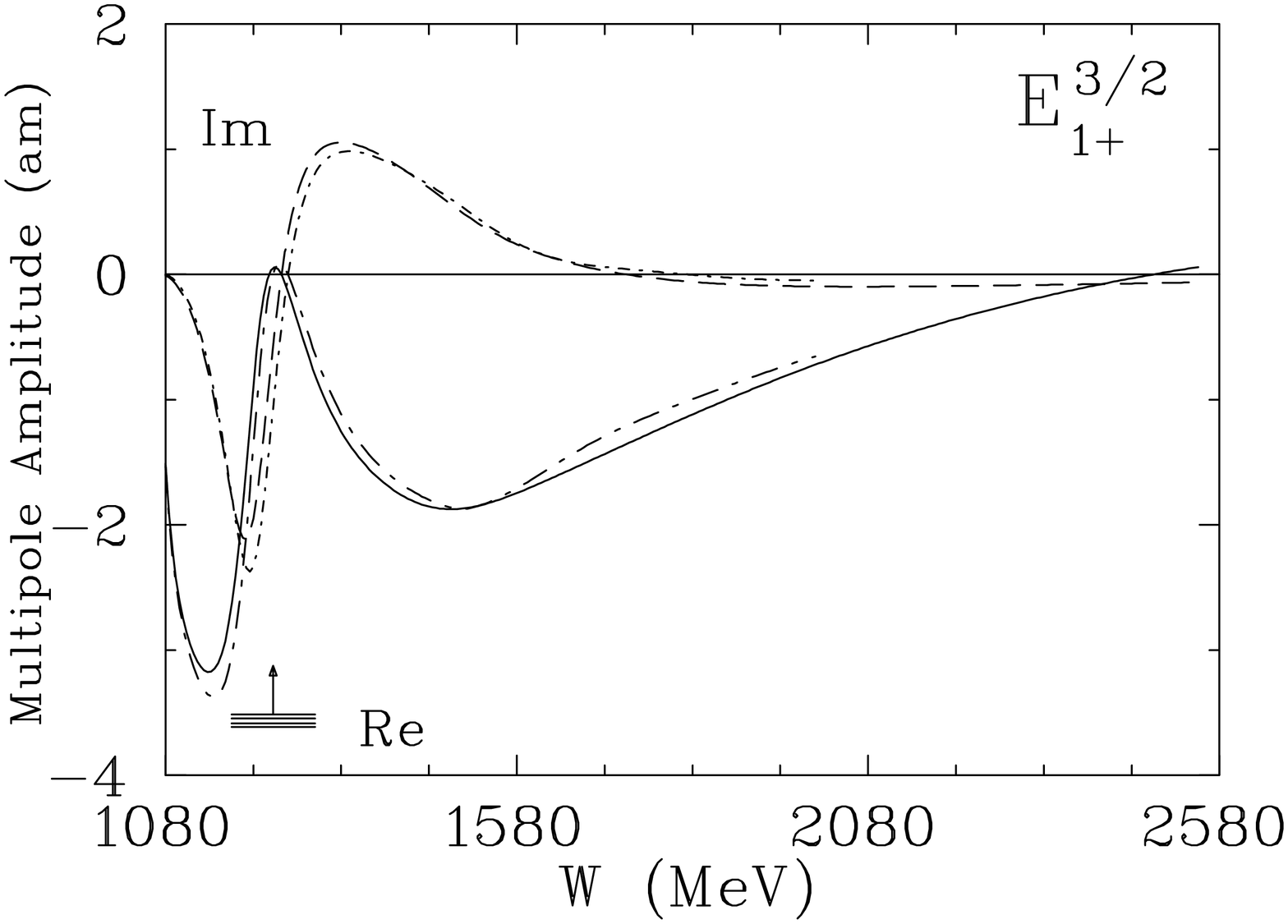}\hfill
\includegraphics[height=0.45\textwidth, angle=90]{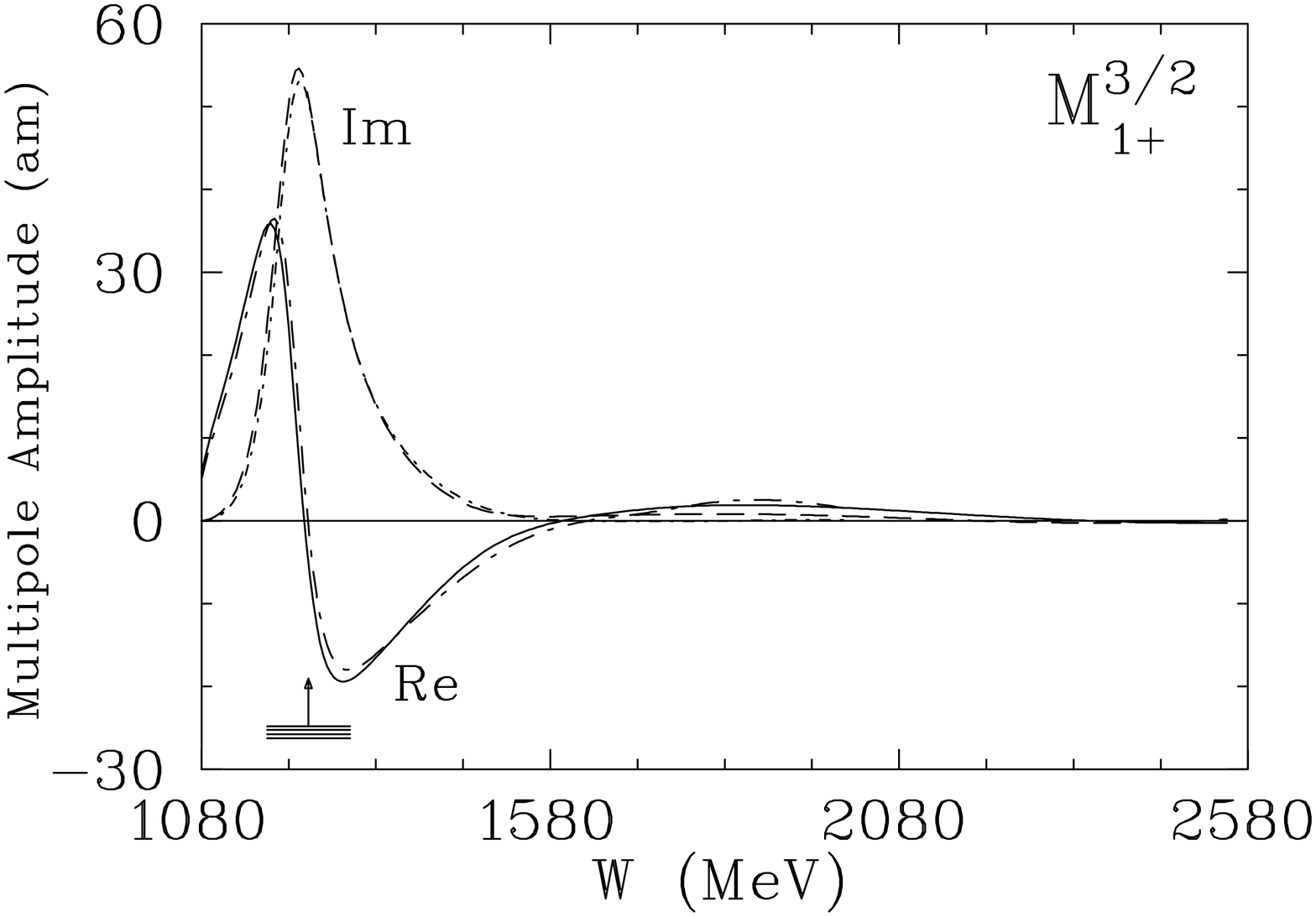}}
\centerline{
\includegraphics[height=0.45\textwidth, angle=90]{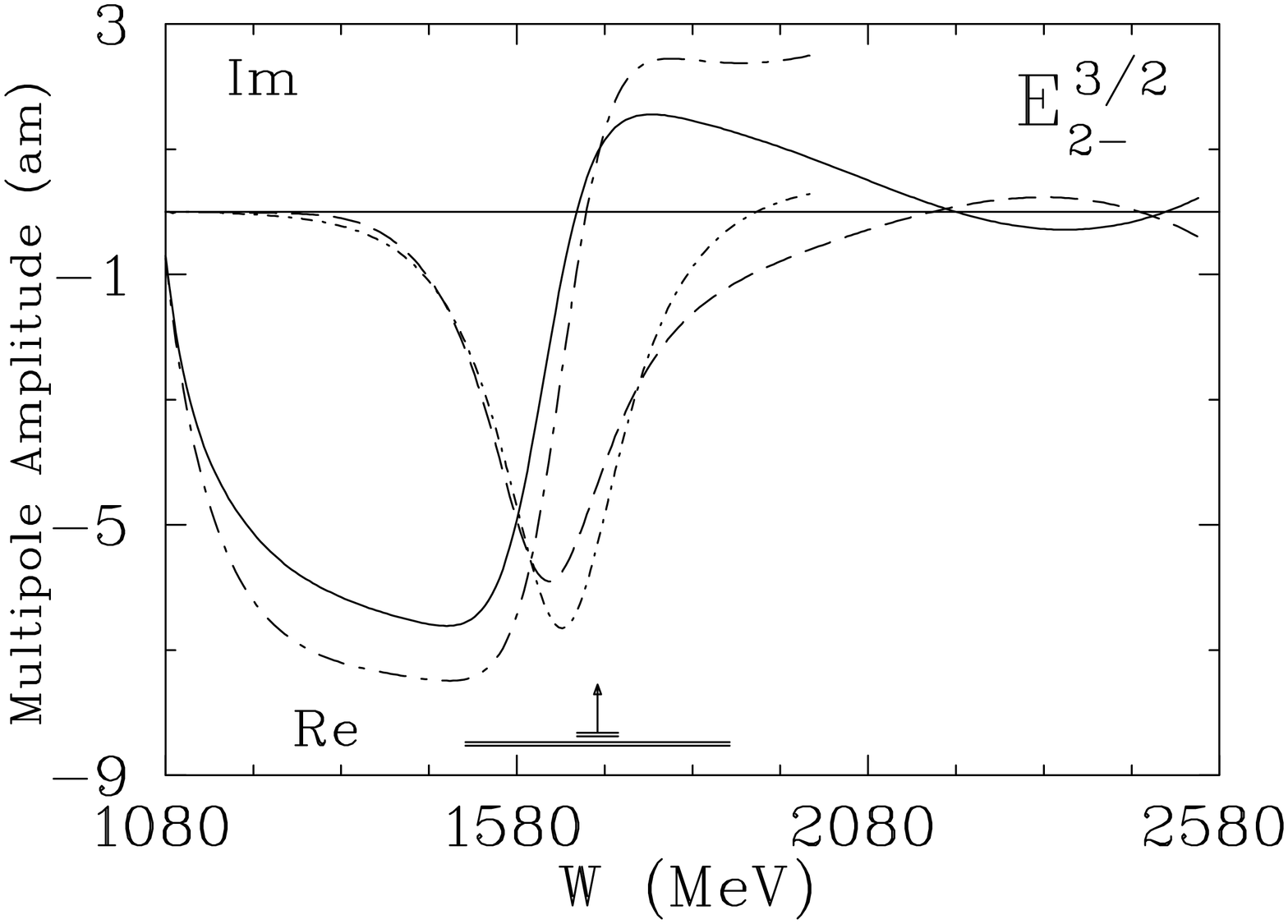}\hfill
\includegraphics[height=0.45\textwidth, angle=90]{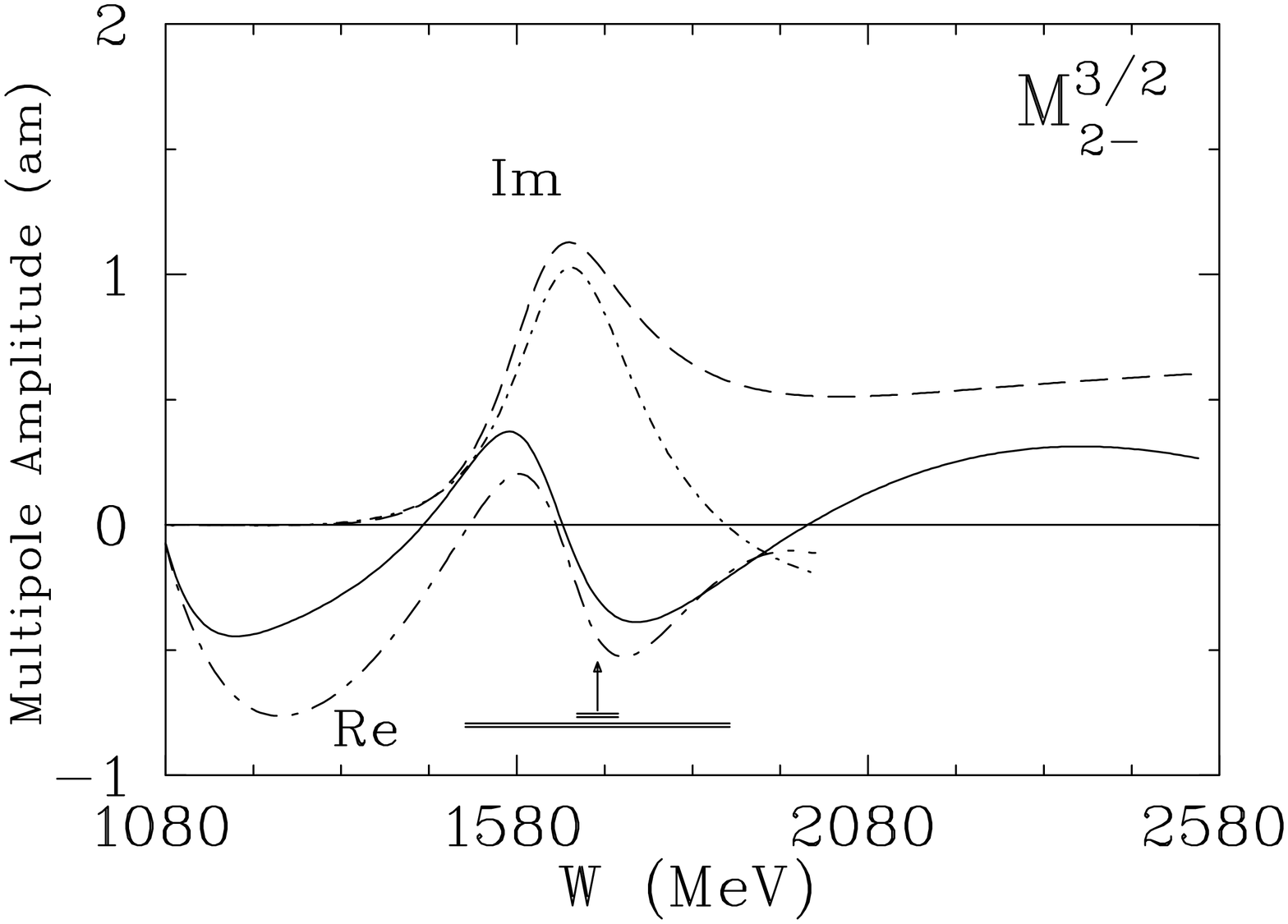}}
\centerline{
\includegraphics[height=0.45\textwidth, angle=90]{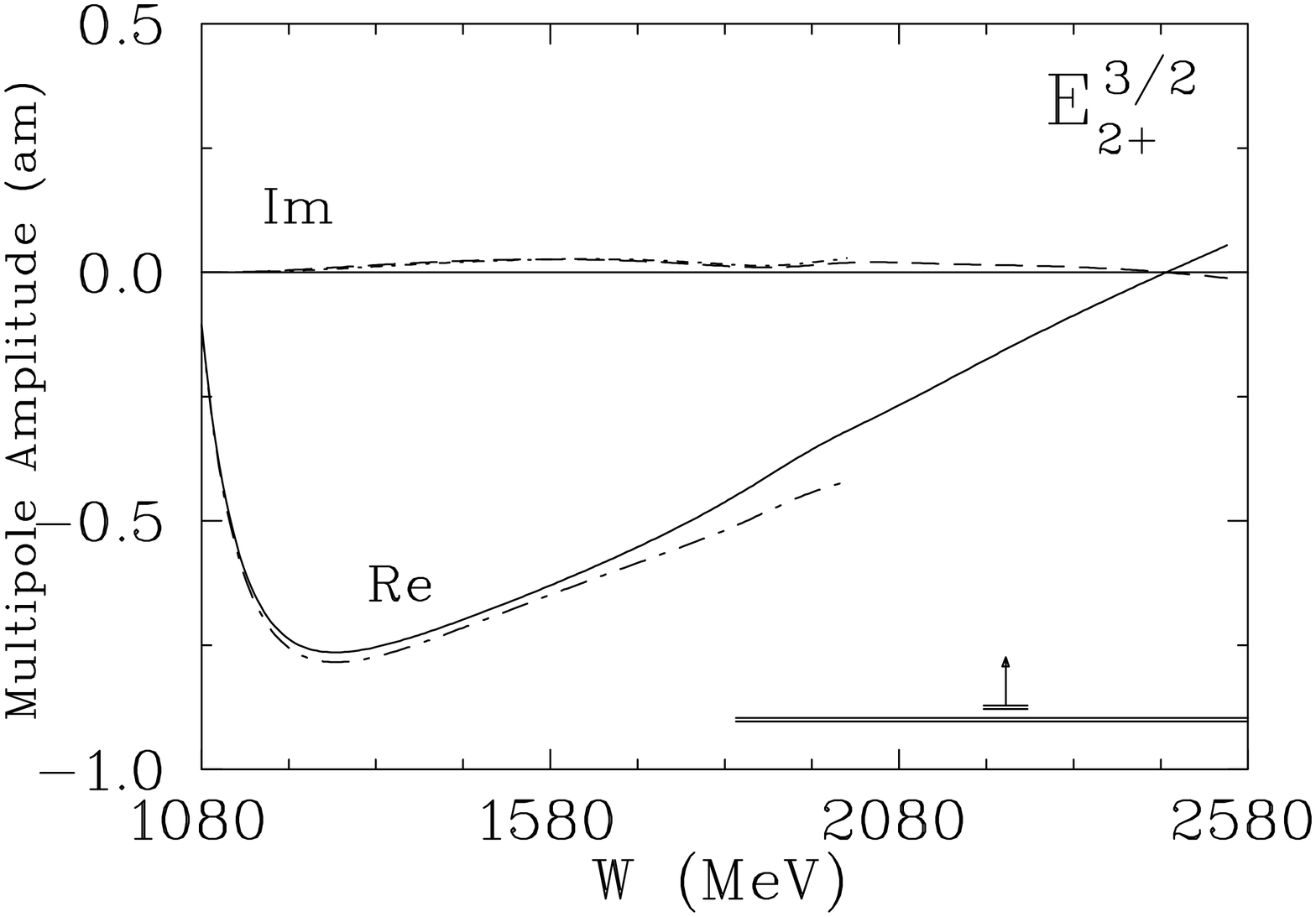}\hfill
\includegraphics[height=0.45\textwidth, angle=90]{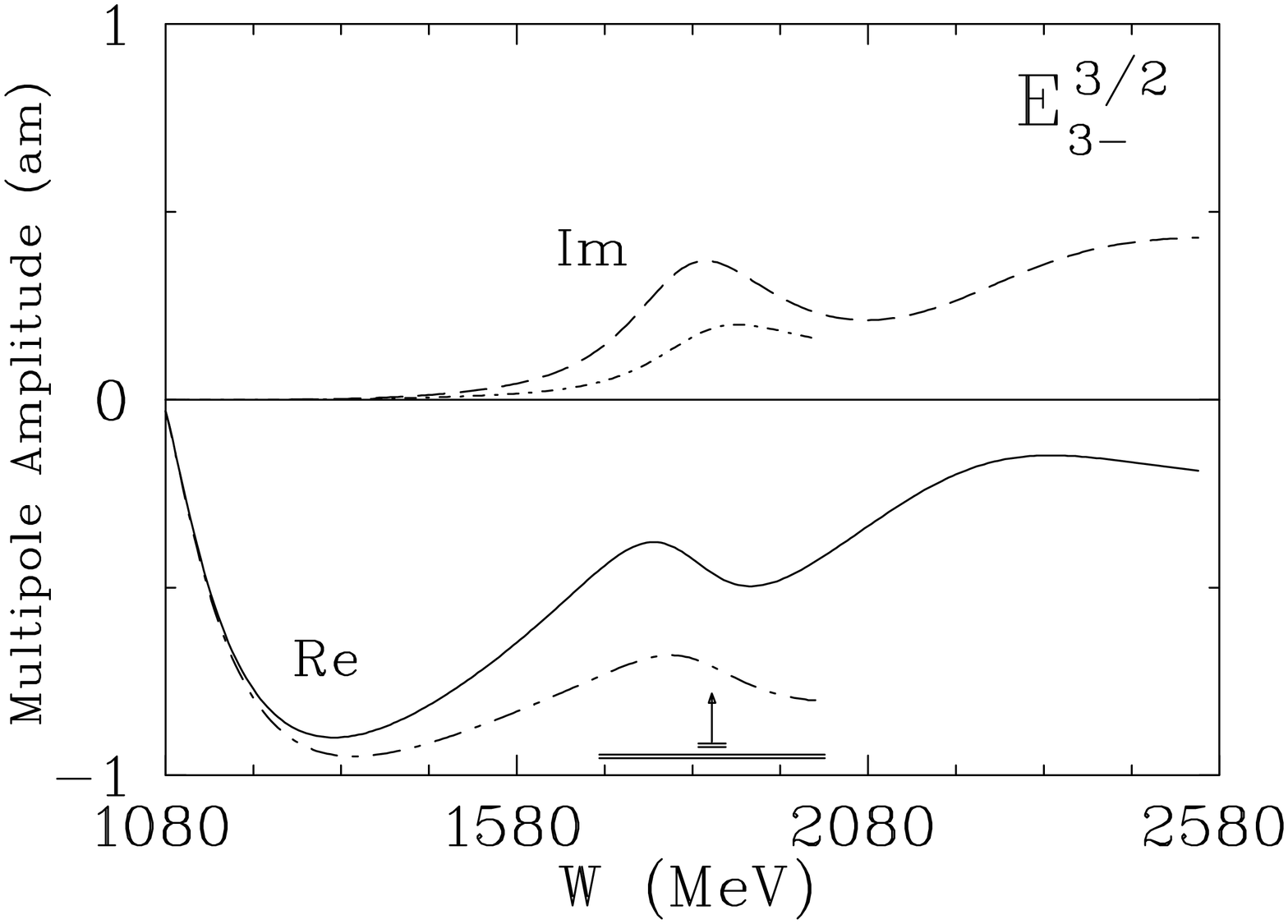}}
\caption{Multipole amplitudes from threshold to $E_{\gamma}$ 
         = 3~GeV for isospin $3/2$.  Notation as in 
         Fig.~\protect\ref{fig:g6}. \label{fig:g7}}
\end{figure*}

Multipoles from the FA06 fit are compared to the earlier SAID (SM02)
determinations in Figs.~\ref{fig:g4} and ~\ref{fig:g5}.  As suggested
by the falling cross sections, the multipoles are either flat or
rapidly decreasing in magnitude at the upper energy limit (e.g., the
$E^{1/2}_{2-}$ and $E^{3/2}_{1+}$ multipoles).  The extension to
higher energies provided by this data set has resulting in a smoothing
of some structures found in the 2~GeV limit of the SM02 solution (see,
in particular, the $E^{3/2}_{2-}$ and $E^{3/2}_{2+}$ multipoles).
Comparisons with the MAID05 solution are given in Figs.~\ref{fig:g6}
and~\ref{fig:g7}.

Resonance contributions have been fit and the resulting helicity
amplitudes are presented in Table~\ref{tab:tbl3}.  Values for the
resonance mass $W_{R}$, width $\Gamma$, and branching fraction
$(\Gamma_{\pi N} / \Gamma)$ for the various resonances were taken from
the recent SAID analysis of pion-nucleon elastic
scattering~\cite{sp06}.  The electromagnetic resonance couplings were
extracted using a form
\begin{equation}
B(W)(1 + i T_{\pi N} ) + T_{BW}e^{i\phi},
\end{equation}
wherein $T_{\pi N}$ was the associated full pion-nucleon $T$-matrix
and $T_{BW}$ was a Breit-Wigner parametrization of the resonance
contribution. This is similar to the MAID form, but allows for a more
flexible polynomial function to account for the non-resonant
contributions from channels such as $\pi\Delta$, as described in
Ref.~\cite{ron}.

\begin{table}[th]
\caption{Resonance parameters for N$^\ast$ and $\Delta^\ast$
         from the SAID fit to the $\pi N$ data ~\protect\cite{sp06}
         (second column) and helicity amplitudes $A_{1/2}$ and
         $A_{3/2}$ (in [(GeV)$^{-1/2}*10^{-3}$] units) from the FA06
         solution (first row) and average values from the
         PDG06~\protect\cite{PDG} (second row). \label{tab:tbl3}}
\vspace{2mm}
\begin{tabular}{|c|c|c|c|}
\colrule
Resonance        & $\pi N$ SAID               &   $A_{1/2}$    & $A_{3/2}$ \\
\colrule
$N(1535)S_{11}$  & $W_{R}$=1547~MeV           &   91.0$\pm$2.2 & \\
                 & $\Gamma$=188~MeV           &   90$\pm$30    & \\
                 & $\Gamma _{\pi}/\Gamma$=0.36&                & \\
$N(1650)S_{11}$  & $W_{R}$=1635~MeV           &   22.2$\pm$7.2 & \\
                 & $\Gamma$=115~MeV           &   53$\pm$16    & \\
                 & $\Gamma _{\pi}/\Gamma$=1.00&                & \\
$N(1440)P_{11}$  & $W_{R}$=1485~MeV           &$-$50.6$\pm$1.9 & \\
                 & $\Gamma$=284~MeV           &$-$65$\pm$4     & \\
                 & $\Gamma _{\pi}/\Gamma$=0.79&                & \\
$N(1720)P_{13}$  & $W_{R}$=1764~MeV           &   96.6$\pm$3.4 & $-$39.0$\pm$3.2 \\
                 & $\Gamma$=210~MeV           &   18$\pm$30    & $-$19$\pm$20    \\
                 & $\Gamma _{\pi}/\Gamma$=0.09&                & \\
$N(1520)D_{13}$  & $W_{R}$=1515~MeV           &$-$28.0$\pm$1.9 & 143.1$\pm$2.0   \\
                 & $\Gamma$=104~MeV           &$-$24$\pm$9     & 166$\pm$5       \\
                 & $\Gamma _{\pi}/\Gamma$=0.63&                & \\
$N(1675)D_{15}$  & $W_{R}$=1674~MeV           &   18.0$\pm$2.3 &  21.2$\pm$1.4   \\
                 & $\Gamma$=147~MeV           &   19$\pm$8     &  15$\pm$9       \\
                 & $\Gamma _{\pi}/\Gamma$=0.39&                & \\
$N(1680)F_{15}$  & $W_{R}$=1680~MeV           &$-$17.3$\pm$1.4 & 133.6$\pm$1.6   \\
                 & $\Gamma$=128~MeV           &$-$15$\pm$6     & 133$\pm$12      \\
                 & $\Gamma _{\pi}/\Gamma$=0.70&                & \\
\colrule
\colrule
$\Delta(1620)S_{31}$& $W_{R}$=1615~MeV        &   49.6$\pm$2.2 & \\
                 & $\Gamma$=147~MeV           &   27$\pm$11    & \\
                 & $\Gamma _{\pi}/\Gamma$=0.32&                & \\
$\Delta(1232)P_{33}$& $W_{R}$=1233~MeV        &$-$139.1$\pm$3.6& $-$257.6$\pm$4.6\\
                 & $\Gamma$=119~MeV           &$-$135$\pm$6    & $-$250$\pm$8    \\
                 & $\Gamma _{\pi}/\Gamma$=1.00&                & \\
$\Delta(1700)D_{33}$& $W_{R}$=1695~MeV        &  125.4$\pm$3.0 & 105.0$\pm$3.2   \\
                 & $\Gamma$=376~MeV           &  104$\pm$15    &  85$\pm$22      \\
                 & $\Gamma _{\pi}/\Gamma$=0.16&                & \\
$\Delta(1905)F_{35}$& $W_{R}$=1858~MeV        &   21.3$\pm$3.6 & $-$45.6$\pm$4.7 \\
                 & $\Gamma$=321~MeV           &   26$\pm$11    & $-$45$\pm$20    \\
                 & $\Gamma _{\pi}/\Gamma$=0.12&                & \\
\colrule
\end{tabular}
\end{table}

The uncertainties for the helicity amplitudes given in
Table~\ref{tab:tbl3} extracted in the FA06 analysis correspond to a
$\chi^2$ increase of 9 in the fit. This uncertainty is thus purely
statistical and does not account for subjectivity in the resonance
extraction and selection of the energy range used in the resonance fit.
Table~\ref{tab:tbl4} compares $\chi^2$ values from FA06 and the
resonance fits over corresponding energy ranges.

\begin{table}[h]
\caption{Comparison of the resonance fit and FA06 against data
         for the isospin $1/2$ and $3/2$ baryon resonances (see text
         and associated Table~\protect\ref{tab:tbl3}). The ``Data"
         corresponds to the number of experimental data points in 
         the given range of the $W_{min}$ and $W_{max}$ limits.
         Energy limits are those used in the most recent $\pi N$
         resonance extractions~\protect\cite{sp06}.
\vspace{2mm}
         \label{tab:tbl4}}
\begin{tabular}{|c|c|c|c|c|c|}
\colrule
Resonance      & $W_{min}$ & $W_{max}$ & Fit      & FA06     & Data\\
               & (MeV)& (MeV)& $\chi^2$ & $\chi^2$ &     \\
\colrule
$N(1535)S_{11}$& 1490 & 1590 & 7129     & 7546     & 3552\\
$N(1650)S_{11}$& 1620 & 1770 & 6013     & 6520     & 2588\\
$N(1440)P_{11}$& 1350 & 1550 &13434     &13752     & 6897\\
$N(1720)P_{13}$& 1650 & 1790 & 5520     & 5717     & 2514\\
$N(1520)D_{13}$& 1480 & 1560 & 7252     & 7554     & 3481\\
$N(1675)D_{15}$& 1610 & 1730 & 5900     & 6139     & 2406\\
$N(1680)F_{15}$& 1620 & 1730 & 5359     & 5625     & 2187\\
\colrule
\colrule
$\Delta(1620)S_{31}$& 1570 & 1680 & 5764     & 6018     & 2475\\
$\Delta(1232)P_{33}$& 1180 & 1270 & 6303     & 6630     & 3351\\
$\Delta(1700)D_{33}$& 1550 & 1750 & 7285     & 7577     & 3064\\
$\Delta(1905)F_{35}$& 1770 & 1920 & 3930     & 4089     & 1787\\
\colrule
\end{tabular}
\end{table}

Most of the resonance couplings determined in this analysis generally
are in fair agreement with the PDG averages, but there are significant
disagreements for the $N(1650)$ and $N(1720)$. The $N(1650)$ couplings
are particularly difficult to extract as there is an overlapping
resonance (the $N(1535)$ resonance) below this state and possibly a
third resonance slightly higher in energy. The $N(1720)$ coupling,
quoted by the PDG, is not clearly determined even in
sign. Furthermore, the multipoles associated with this state have also
changed dramatically in the extension to 3~GeV, as can be seen in
Fig.~\ref{fig:g4}.

The good agreement between the fit and PDG parameters seen for the
$N(1535)$ resonance also deserves some comment. The large PDG error
band was given mainly to account for the spread in determinations from
pion and eta photoproduction analyses. This discrepancy has largely
disappeared~\cite{n1535}.  However, the same qualifications addressed
to the $N(1650)$ resonance apply here as well. Model dependence in
this extraction is certainly larger than the statistical error quoted
in Table~\ref{tab:tbl3}.

Given the smooth behavior exhibited by the excitation functions in
Figs.~\ref{fig:g3} and ~\ref{fig:g3a}, the CLAS cross sections provide
no hint of ``missing" resonance structure between 2 and 3~GeV.  The
SAID fits implicitly contain only those resonances found in the
corresponding SAID analysis of elastic pion-nucleon scattering data.
No change in the form of the SAID photoproduction fit was found to be
necessary. In contrast, the CB-ELSA fit required many additional
resonance contributions, some of which are 1- and 2-star rated PDG
states, as well as a new N(2070) resonance. One possible explanation
is apparent in Fig.~\ref{fig:g3a}, which shows the CLAS data to be
somewhat smoother than the CB--ELSA excitation functions.
Model-dependence in the separation of resonance and background
contributions is also a critical factor.  This uncertainty can be
reduced through measurements of further (polarized) data.


\section{Conclusion}
\label{sec:conc}

The differential cross sections for the reaction $\piZeroRxn$ for
incident photon energies between 0.675 and 2.875~GeV have been
measured.  
These measurements have been compared to existing data, mainly from the
CB-ELSA Collaboration. The overall agreement is excellent, though
statistically significant disagreement is evident in some cases between
that data and a few values obtained here in the forward direction.
The variation of fits in the forward region suggest further
measurements at those angles would be very useful. As shown in
Figs.~\ref{fig:g3} and ~\ref{fig:g3a}, both the SAID fit SM02,
developed prior to the publication of both the CB-ELSA and CLAS data,
and FA06 (including both sets) appear to favor the most forward CLAS
measurements over the CB-ELSA measurements.

Multipole analyses incorporating the CLAS data and extending to 3~GeV
now appear more smooth near the 2~GeV limit of previous fits, tending
to be relatively featureless and rapidly decreasing above this
energy. At energies above 2~GeV and outside of the angular range
constrained by CLAS data, the fits were found to be unstable. More
progress will require precise measurements in that angular region, and
also the availability of polarization observables of a similar quality
and coverage. Corresponding data from $n \pi^+$ photoproduction are
also clearly required to make an isospin decomposition.  
A partial wave analysis performed in this work using only ``4-star''
resonances at or below $W$=1910 MeV satisfactorily described the data.

\acknowledgments

The authors gratefully acknowledge the Jefferson Lab Accelerator
Division staff.  This work was supported by the National Science
Foundation, the Department of Energy (DOE), the Deutsche
Forschungsgemeinschaft (through an Emmy Noether grant to U.T.), the
French Centre National de la Recherche Scientifique and Commissariat
\`a l'Energie Atomique, the Italian Istituto Nazionale di Fisica
Nucleare, and the Korean Science and Engineering Foundation.  The
Southeastern Universities Research Association (SURA) operated
Jefferson Lab for DOE under contract DE-AC05-84ER40150 during this
work.

\clearpage



\begin{thebibliography}{0}
\bibitem{n1535}This discrepancy was already absent in our
               previous analysis~\protect\cite{sm02} but the
               published value was incorrect due to an error in
               the SAID code.
\end{thebibliography}
\end{document}